\documentclass{aa}  

    \usepackage{graphicx}
    \usepackage{txfonts}
    \usepackage{comment}
    \usepackage{stackengine}
    \usepackage{threeparttable}
    \usepackage[colorlinks = true, allcolors=blue]{hyperref}
    \newcommand{\vik}[1]{{\color{red} #1}}

    \usepackage[normalem]{ulem}
    \usepackage{stfloats}
    \usepackage{booktabs}
    \usepackage{orcidlink}
    \usepackage{sidecap}
    \usepackage{pdflscape}
    \usepackage{amsmath}
    \usepackage{lscape}
    \usepackage{hyperref}
    \usepackage[a4paper]{geometry}
    \newcommand\arcdeg{\mbox{$^\circ$}}%

    \defcitealias{2024arXiv240305143H}{HR2024}
    \defcitealias{2019A&A...627A...4Roser_Praesepe}{RS2019}
\begin{document}

       \title{Tidal tails of nearby open clusters}
       \subtitle{I. Mapping with Gaia DR3}
       \titlerunning{Tidal tails of open clusters}
       \authorrunning{Risbud, Jadhav and Kroupa}
       
       \author{Dhanraj Risbud\orcidlink{0000-0001-9174-2883}\inst{1},
               Vikrant V. Jadhav\orcidlink{0000-0002-8672-3300}\inst{2},
               Pavel Kroupa\orcidlink{0000-0002-7301-3377}\inst{2,3}
              }
    
       \institute{
            Rheinische Friedrich-Wilhelms-Universität Bonn, Regina-Pacis-Weg 3, D-53113 Bonn, Germany \\ \email{s14drisb@uni-bonn.de}
            \and
            Helmholtz-Institut für Strahlen- und Kernphysik, Universität Bonn, Nussallee 14-16, D-53115 Bonn, Germany\\
            \email{vjadhav@uni-bonn.de}
            \and
            Astronomical Institute, Faculty of Mathematics and Physics, Charles University, V Holešovičkách 2, CZ-180 00 Praha 8, Czech Republic
            }
    
       \date{Received January 13, 2023; accepted June 27, 2023}
    
    % \abstract{}{}{}{}{} 
    % 5 {} token are mandatory
     
      \abstract
      % context heading (optional)
       {
       Tidal tails of open clusters are the result of stellar evaporation from the cluster through the Galactic potential and internal dynamics. With the recent availability of high-precision data, tidal tails are being detected for most of the nearby open clusters.
       }
      % aims heading (mandatory)
       {
       We identify the tidal tail members for all open clusters within a distance of 400 pc that are older than 100 Myr and have $>$100 members. To do this, we use model-independent methods. 
       %Additionally, we aim to identify the biases in the identification method using $N$-body simulations.
       }
      % methods heading (mandatory)
       {
       We used the convergent-point (CP) method to identify the co-moving stars near the open clusters using \textit{Gaia} DR3 data. A new method called the self-compact convergent-point (SCCP) method was proposed and applied to some of the clusters. It performed better overall in tracing the tails. We also analysed the colour-magnitude diagrams and orbital energy to diagnose possible contamination.
       }
      % results heading (mandatory)
       {
       Nineteen out of 21 clusters have tidal tails. Five of them were discovered for the first time through this work. The typical span of the tidal tails is 20--200 pc, and 30--700 member stars lie in the region inside the tidal radius and the tidal tails. Four out of 19 tidal tails are tilted away from direction of the Galactic centre. This contradicts the known theory of the tidal-tail formation. The luminosity functions of the tails and clusters are consistent with each other and with the canonical stellar interstellar mass function, but systematically higher radial velocities for the trailing tail than for the leading tail were observed for the first time. 
       }
      % conclusions heading (optional)
       {
       The CP method is useful for detecting tidal tails on a scale of $\approx$100 pc for clusters closer than 400 pc. A further analysis of theoretical $N$-body models is required to understand the incompleteness and biases in the current sample of tidal tails.
       }
    
    
       \maketitle
    %
    %-------------------------------------------------------------------
    
\section{Introduction} \label{sec:introduction}
% Abbreviations: OC, MW, GC
% regex: \b(?:[A-Z]){2,}

Stars escape from a star cluster potential because the %various dynamical evolution processes concerned with the%
energy equipartition becomes unbound by the tidal field of the host galaxy. These escaped stars form tidal tails around the star cluster in the Galactic gravitational field. These tidal tails of clusters were confidently predicted in the 1940s by \citet{1934HarCi.384....1B} and \citet{1940MNRAS.100..396S}, based upon the very preliminary classical mechanical equations of motion for cluster stars that move in the intra-cluster and Galactic potential. The authors used the best known model of the Milky Way (MW) in those early years.
With modern observing facilities in the 1990s and later, several new discoveries provided observational confirmation of these extended structures around globular clusters ( \citealt{1995AJ....109.2553Grillmair, 1997A&A...320..776Lehmann, kharchenko1997membership, 2001ApJ...548L.165Odenkirchen, 2002AJ....124..349Rockosi, 2006ApJ...637L..29Belokurov, lee2003wide, ibata2019streams}) and on extragalactic scales \citep{2007A&A...473..429Federici}. The Galactic globular clusters are very massive with significant overdensities in five- or six dimensional phase space. They can therefore be fairly easily distinguished from Galactic background stars because they reside in the Galactic halo, which has a low background stellar density. Unlike for globular clusters, it is intrinsically and observationally difficult to map tidal tails in open clusters (OCs). Most of the OCs are found in the Galactic disk close to the Galactic plane, where the noise due to the field stars is most dominant. Their masses are lower and there are fewer stars in a typical OC, and this hampers a detection of the overdensities above the Galactic noise in any phase space. These studies therefore only became possible with the advent of the Gaia mission \citep{2016A&A...595A...1_Gaia_Instrument,2018A&A...616A...1Gaia_DR2_Cit,2021_Gaia_EDR_3,Gaia2023A&A...674A...1G}, which provides accurate positions, parallaxes, and proper motions, for nearly 1.5 billion stars within the MW. 

The recent discoveries and detailed studies of tidal tails of OCs include several different methods that cover several dozen OCs with known tails. Hyades is unarguably the most extensively studied OC with tidal tails. It was mapped by \citet{2019A&A...621L...3Meingast}, \citet{2019A&A...621L...2Roser_Hyades}, and \citet{2021A&A...647A.137Jerabkova} using data from Gaia DR2. \citet{2019A&A...621L...3Meingast} relied on the three-dimensional velocity distribution and spatial distribution of stars constructed from six coordinates, and selected highly comoving stars with respect to the cluster. This revealed a structure for 238 stars with a long extent of 200 pc and a thickness of only 25 pc that covered over 100 deg on the sky, with a typical \reflectbox{S}-shape
\footnote{The S Shape of tidal tails from semi-analytic modelling by \citet{2020A&A...640A..84Dinnbier} predicts that the leading tail has a slight tilt towards the Galactic Center which gives an S-shape in the coordinate system in which the $X-$axis points towards the Galactic Anti-centre, as assumed by the authors. Throughout this work we assumed the coordinate system described in the Section~\ref{sec:finding_overdensity} with the $X-$axis pointing towards the Galactic centre, hence the shape of tails translates to a mirror image, \reflectbox{S}.} 
in the galactocentric $X-Y$ plane, in agreement with the theoretical models \citep{2020A&A...640A..84Dinnbier, 2020MNRAS.497..536Wang}. \citet{2019A&A...621L...2Roser_Hyades} adopted the CP method from \citet{2009A&A...497..209V} and constructed a five-dimensional phase space, where the cluster and tails showed significant overdensities.  Overall, the structure had an extent of around 170 pc in the leading and 70 pc in the trailing tail directions, with 972 stars in total in the cluster and tails. The authors also reported a distinct moving group within the tidal tails. \citet{2021A&A...647A.137Jerabkova} developed N-body models of a Hyades-like cluster, and introduced the new compact convergent-point (CCP) method with which they identified the Hyades tidal tail stars. This work with Gaia DR2 and eDR3 revealed that the tails extend to almost 1 kpc. The authors also noted that for a Hyades-like cluster with an age of 600--700 Myr, the proper motions of the tail stars can differ by up to $\pm$40 km s$^{-1}$ relative to the cluster centre, and they are therefore not simple overdensities in any parameter space.

\citet{2021A&A...645A..84Meingast} used the CP method to find velocities in phase space and DBSCAN to find phase-space overdensities. They identified tidal tails of 10 OCs near the Sun within an age range of 30-300 Myr. The structures around the compact cluster core, as reported by the authors, extend farther than $\geq$ 100 pc and comprise the bulk of the stellar mass. They are comoving with the cluster, but have asymmetric velocity distributions. \citet{10.1093/mnras/stac2906} used several machine-learning algorithms and detected extended structures outside the tidal radius around the targeted 46 OCs with elongated morphologies, 20 of which were reported to show signs that the tidal tails were aligned with the orbit of the cluster in galactocentric coordinates. The limitations arising from the method they used truncated the tails at very short distances. The shortest tails ended a few parsecs outside the tidal radius. \citet{2024arXiv240618767Kos} simulated the dissolution of OCs and detected tidal tails in 476 OCs with a probabilistic method. 

\citet{2023A&A...679A.105Vaher} developed a new method for tracing the escaped stars from the nearby clusters and applied it to the same to ten OCs near the Sun. Using the radial velocities from the Gaia data as well, the authors performed trace-back computations to determine probabilistic estimates whether an escaped star was a previous member of a cluster. Because the subsample with radial velocities was limited, the number of escaped stars for every cluster they found was smaller than with other dedicated methods for tracing the tidal tails. At the same time, they were able to trace the stars a few hundred parsec away from the clusters. Other notable attempts to detect tidal tails of OCs include Coma Berenices / Melotte 111 \citep{2019A&A...624L..11Furnkranz, 2019ApJ...877...12Tang, 2021ApJ...912..162Pang}, Ruprecht 147 \citep{2019AJ....157..115Yeh}, Praesepe/ NGC 2632 \citep{2019A&A...627A...4Roser_Praesepe}, NGC 2506 \citep{2020ApJ...894...48Gao}, UBC 274 
\citep{2020A&A...635A..45Castro-Ginard}, Blanco 1 \citep{2020ApJ...889...99Zhang,2021ApJ...912..162Pang,2023A&A...679A.105Vaher}, Alpha Persi \citep{2020AJ....160..142Nikiforova, 2021A&A...645A..84Meingast}, NGC 2516, NGC 6633 \citep{2021ApJ...912..162Pang}, Pleiades / Melotte 22 \citep{2021RNAAS...5..173Li,2021A&A...645A..84Meingast,2023A&A...679A.105Vaher} , IC 4756 \citep{2021AJ....162..171Ye}, and NGC 752 \citep{2021MNRAS.505.1607Bhattacharya, 2022MNRAS.514.3579Boffin}.

The shapes of tidal tails are influenced by many known and unknown variables on local as well as on Galactic scales. A qualitative explanation for the formation of leading and trailing tails stems from the differential velocities of stars that leave the OC, as explained by \citet{LectureNotesKroupa}. Stars with a lower and higher Galactic velocity than the cluster form leading and trailing tails, respectively, with the typical S-shape. This was further discussed in detail in the semi-analytic model by \citet{2020A&A...640A..84Dinnbier} and \citet{2021A&A...647A.137Jerabkova}. \citet{2020A&A...640A..84Dinnbier} also made the distinction into two types of tidal tails on the basis of the mechanism that is dominant in taking out the stars from the cluster into the tails and based on the timescales over which the tails form. Type I tails are formed relatively quickly by stars that escape from the cluster just after gas expulsion, and type II tidal tails form over hundreds of million years through the gradual evaporation of stars through energy equipartion. The morphology and kinematics of these two types are significantly different and were previously studied in detail by \citet{2008MNRAS.387.1248Kupper,2010MNRAS.401..105K} for type  II and by \citet{2020A&A...640A..84Dinnbier} for type I tidal tails.

Another interesting phenomenon of tidal tails includes substructures that are called epicyclic overdensities or K{\"u}pper overdensities. These are density enhancements along the tidal tails at specific locations for a cluster in both circular and eccentric orbits \citep{2008MNRAS.387.1248Kupper,2010MNRAS.401..105K,2009MNRAS.392..969Just}. For a cluster on a circular orbit around the Galactic centre, the distance of an overdensity from the cluster centre depends on the mass of the cluster and on the strength of the tidal field. It is a decreasing function with time as the cluster looses its mass \citep{2010MNRAS.401..105K}. More complexity arises for clusters with eccentric orbits around the Galactic centre, where the distance continuously oscillates with time \citep{2010MNRAS.401..105K}. 

As most of the OCs reside in the Galactic disk, the shapes of their tidal tails are influenced by the gravitational potential of the inner Galaxy. This provides possible probes for measuring the potential. Hence, the sample of tidal tails of OCs nearby the Sun, which shows wide ranges in different physical properties such as mass, ages, and orbital parameters, is very important for understanding the local gravitational potential. Further, as pointed out by \citet{2021A&A...655A..71Wang}, tidal tails are important tools for studying the birth condition and evolution of star clusters and of the coupling and interaction of the clusters with the Galactic potential. The orientation of tidal tails with respect to the velocity vector of the cluster and current angular momentum of the cluster can also constrain the rotation of the cluster \citep{2021A&A...647A.137Jerabkova, 2023A&A...673A.128Garcia}. 
The K{\"u}pper overdensities mentioned above also provide an additional constraint on the Galactic potential \citep{2015ApJ...803...80Kupper}.
The stellar content of a cluster that is caused by the stellar evolution is irrecoverably lost, impacting the initial mass function, but the expelled stellar content residing in the tidal tails holds information that leads to a better completeness when it is accounted for \citep{2024A&A...691A.227Wirth}. Finally, the number of stars in the leading tail versus the number of stars in the trailing tail allows us to test gravitation theories \citep{2022MNRAS.517.3613Kroupa,2024ApJ...970...94Kroupa,2024arXiv241113675Pflamm}. 

The aim of this work was to detect the tidal tails of nearby OCs. The structure of the paper is as follows: The Gaia DR3 data and cluster sample selections are explained in Section~\ref{sec:data}.
Section~\ref{sec:methods} describes CP-based tail detection methods, and we present the results in Section~\ref{sec:results}. Further, we discuss the tidal tail properties and possible incompleteness in Section~\ref{sec:discussion}. Section~\ref{sec:conclusions} summarises the findings in this work. The auxiliary materials can be found in Appendix~\ref{sec:Fig_Desc}.

\section{Data} \label{sec:data}
\subsection{Gaia data} \label{Sec:Gaia_Data}
We made use of Gaia DR3 \citep{Gaia2023A&A...674A...1G}, which provides accurate and precise astrometry and photometry in three broad-band Gaia filters (G, BP, and RP) for nearly 1.5 billion sources. We downloaded data for all the sources with parallax $\omega$ > 2 mas. Quality cuts were applied to include stars with \textsc{astrometric\_excess\_noise} < 1 or  \textsc{astrometric\_excess\_noise\_sig} < 2 and \textsc{ruwe} < 1.4. Additionally, we calculated the absolute distance error for every source with an error propagation as follows:
\begin{equation}\label{eq:Dist_err}
\sigma_{\text{Dist}} = \frac{1000\sigma_{\omega}}{\omega_{2}} \quad [pc],
\end{equation}
where $\sigma_{\omega}$ is the error in the parallax in mas yr$^{-1}$. 
We constrained $\sigma_{\text{Dist}}$ < 10 pc. This modification was made to rule out any parallax-introduced bias that may occur in the distribution of stars in the leading and trailing tails as a result of their orientation. The ADQL query is given in Section~\ref{sec:adql}. The final sample of stars with distances up to 500 pc we obtained after these cuts constitutes 7,723,781 individual sources.

\subsection{Sample selection}

We selected nearby older OCs from \citet{2024arXiv240305143H} (\citetalias{2024arXiv240305143H} hereafter) within 400 pc and with an age greater than 100 Myr with a minimum of 100 stars within the tidal radius according to Table \ref{Table: Clusters_basics_params}. The criterion for the minimum number of stars was placed to ensure that the CP of the cluster can be reliably found and that the region inside the tidal radius of the cluster is well traced by the neighbour-finding algorithm (see Section~\ref{sec:finding_overdensity}). This selection yielded a sample set of a total of 22 OCs, 21 of which were selected for this work\footnote{We did not include  
Herschel 1 because the CP of this cluster found by the method described in Section~\ref{sec: CP} has large errors. This makes it unreliable for any further analysis. The central stellar concentration of the cluster was found to be relatively low, which makes it difficult for the neighbour-finding algorithm to work (see Section~\ref{sec:finding_overdensity}).}. 
Their known parameters are listed in Table~\ref{Table: Clusters_basics_params}.  

\section{Methods} \label{sec:methods}

\subsection{CP method}\label{sec: CP}

Stars within OCs are known to be spatially co-moving with similar space velocities (and proper motions) that appear to converge at a single point on the celestial sphere. The motion is a perspective effect with respect to the solar barycentric frame of reference, and the point is referred to as the CP. The well-established CP method is a tool for simultaneously determining the CP as well as the actual members of moving groups or OCs \citep{1971MNRAS.152..231J,1999MNRAS.306..381D,2009A&A...497..209V}. As we already had information on cluster priors from \citetalias{2024arXiv240305143H}, we combined and modified the basic CP method \citep{1971MNRAS.152..231J,1999MNRAS.306..381D} to make it best suitable for this sample, for which the cluster members are known. 

We only considered the member stars within the Jacobi radius or tidal radius from \citetalias{2024arXiv240305143H} for every cluster to calculate the CP. We laid a grid of test CPs ranging from -90\arcdeg to +90\arcdeg in Dec. and 0\arcdeg to 180 \arcdeg or 180\arcdeg  to 360\arcdeg in R.A., depending upon the hemisphere of the cluster position on the sky with an initial resolution of 1\arcdeg in R.A. and Dec. For each individual point on the grid, we calculated the observed velocities parallel and perpendicular to the CP, denoted by $V_{\parallel \text{obs}}$,$V_{\perp \text{obs}}$, as follows: 
\begin{equation} \label{eq:v_parallal_obs}
V_{\parallel \text{obs}} = (\sin\theta \, \mu_{\alpha}\cos\delta + \cos\theta \, \mu_{\delta}) \times \frac{4.74047}{\omega} \quad [\rm{km \ s^{-1}}],
\end{equation}

\begin{equation}
V_{\perp \text{obs}} = (-\cos\theta \, \mu_{\alpha}\cos\delta + \sin\theta \, \mu_{\delta} ) \times \frac{4.74047}{\omega} \quad [\rm{km \ s^{-1}}],
\end{equation}
where $\mu_{\alpha}\cos\delta$, $\mu_{\delta}$, and $\omega$ are pmra, pmdec, and the parallax of individual stars from Gaia DR3, respectively. The factor 4.74047 is the ratio of one astronomical unit in kilometers and the number of seconds in one Julian year, which converts proper motions into mas yr$^{-1}$ to velocities in km s$^{-1}$. The position angle $\theta$ of the CP is expressed as
\begin{equation}
\tan{\theta} = \frac{\sin({\alpha_{CP}} - \alpha)}{\cos\delta\tan\delta_{CP} - \sin\delta \cos({\alpha_{CP}} - \alpha)},
\end{equation}
where ($\alpha$, $\delta$) is the on-sky position of individual stars, and ($\alpha\textsubscript{CP}$, $\delta\textsubscript{CP}$) is the position of the test CP from the grid. The predicted velocities were calculated for the entire cluster as follows: 
\begin{equation}
V_{\parallel \text{pred}} = \overline{\mu_{c}} \times\frac{ 4.74047}{\overline{\omega_{c}}} \quad [\rm{km \ s^{-1}}],
\end{equation}
\begin{equation} \label{eq:v_per_pred}
V_{\perp \text{pred}} = 0  \quad [\rm{km \ s^{-1}}],
\end{equation}
where $\overline{\mu\textsubscript{c}}$ denotes the mean proper motion of the cluster, and $\overline{\omega\textsubscript{c}}$ is the mean parallax of the cluster. The equations based on the proper motion from \citet{1999MNRAS.306..381D} were adapted and converted into velocity space in our equations~\ref{eq:v_parallal_obs} to~\ref{eq:v_per_pred}. 
In this formalism, $V_{\parallel \text{obs}}$ corrects for two effects. One effect is the projection effect on the velocities due to the angular distances of individual stars from the CP, and the other effect is the parallax effect. $V_{\perp \text{obs}}$ deals with the intrinsic velocity dispersion of the cluster and measurement errors in the proper motions. We plot the distributions of stars from our set of selected clusters in Figs.~\ref{fig:ASCC_101} to ~\ref{fig:UPK_545} with axes $V_{\parallel \text{obs}} - V_{\parallel \text{pred}} = \Delta V_{\parallel}$  and $V_{\perp \text{obs}} - V_{\perp \text{pred}} = V_{\perp \text{obs}}$. The distributions are centred around (0,0), as expected. Finally, we calculated $X\textsuperscript{2}$ for every test CP as follows:
\begin{equation}
X^{2} = \sum_{i=1}^{N} \frac{\mu_{\perp \text{obs}}}{\sigma_{\perp}} = \sum_{i=1}^{N} \frac{(-\cos\theta \, \mu_{\alpha}\cos\delta + \sin\theta \, \mu_{\delta} )}{\sigma_{\perp}},
\end{equation}
where $N$ is the number of stars we used for the calculation, and $\sigma_{\perp}$ is the first-order error as follows:

\begin{equation}
\sigma_{\perp} = \sqrt{\cos^2(\theta) \cdot \sigma_{\mu_{\alpha}\cos\delta}^2 + \sin^2(\theta) \cdot \sigma_{\mu_{\delta}}^2}.
\end{equation}
The point from the grid with the lowest value of $X\textsuperscript{2}$ was selected as the CP.

The algorithm explained above was bootstrapped to run 100 times with randomly selected 75\% of the stars in each bootstrap loop for each cluster. This yielded a distribution of $\alpha\textsubscript{CP}$ and $\delta\textsubscript{CP}$, which is Gaussian to a good extent, with means $(\overline{\alpha\textsubscript{CP}},\overline{\delta\textsubscript{CP}})$, medians $({\alpha\textsubscript{Med,CP}},{\delta\textsubscript{Med,CP}})$, and standard deviations $(\sigma_{\alpha_{CP}},\sigma_{\delta_{CP}})$. In the next iteration, we reduced the range of test CPs to $({\alpha\textsubscript{Med,CP}} \pm 3\sigma_{\alpha_{CP}},{\delta\textsubscript{Med,CP}} \pm 3\sigma_{\delta{CP}} )$ and boosted the grid resolution to 0.1\arcdeg. The second iteration was run  exactly as the first, and the final mean CPs and the standard deviations for every cluster are listed in Table~\ref{Table: Clusters_basics_params}.

\subsection{Finding phase-space overdensities} \label{sec:finding_overdensity}

We follow the approach presented by \citet{2019A&A...627A...4Roser_Praesepe} (\citetalias{2019A&A...627A...4Roser_Praesepe} hereafter) in this section to reveal the phase-space overdensities for clusters and their possible tidal tails. For all the sources from Section~\ref{sec:data}, we transformed the positions and parallaxes into the Solar System barycentric Galactic Cartesian coordinate system using \textsc{astropy} \citep{astropy:2022}. The X-axis points towards the Galactic centre, the Y-axis points in the direction of rotation of the Galaxy in the Galactic plane orthogonal to the X-axis, and the Z-axis points towards the Galactic north pole, orthogonal to the Galactic plane. We also calculated $\Delta V_{\parallel}$ , $ V_{\perp}$ for every star in the data, given the CPs of the clusters in Table~\ref{Table: Clusters_basics_params}. To reduce the computational burden, a selection of a phase-space volume around every cluster within |Z - Z\textsubscript{c}| $\leq$ 50 pc in the Z plane, |$\Delta V_{\parallel}$| $\leq$ 50 km s$^{-1}$ and |$ V_{\perp}$| $\leq$ 10 km s$^{-1}$ in the velocity plane was made. Another cut in position space out of four of the following: $X$ $\leq$ 0, $Y$ $\leq$ 0, $X$ $\geq$ 0, and $Y$ $\geq$ 0, was made based on the position of the cluster in the chosen coordinate system, selecting only the hemisphere in which the cluster lay. 
% These cuts reduced the sample to be analysed further and were also large enough to include the possible tails of clusters up to a good extent in phase-space. 

The phase-space neighbourhood definition was adopted from \citetalias{2019A&A...627A...4Roser_Praesepe} as follows: Stars $i$ and $j$ with phase-space coordinates $(X\textsubscript{i},Y\textsubscript{i},Z\textsubscript{i},\Delta V_{\parallel\textsubscript{i}},  V_{\perp\textsubscript{i}})$, $(X\textsubscript{j},Y\textsubscript{j},Z\textsubscript{j},\Delta V_{\parallel\textsubscript{j}},  V_{\perp\textsubscript{j}})$ are neighbours if
\begin{equation}\label{eq: space}
(X\textsubscript{i} - X\textsubscript{j})^2 + (Y\textsubscript{i} - Y\textsubscript{j})^2 + (Z\textsubscript{i} - Z\textsubscript{j})^2 \leq r\textsubscript{lim}^2 
\end{equation}
and
\begin{equation}\label{eq: vel}
\frac{(\Delta V_{\parallel\textsubscript{i}} - \Delta V_{\parallel\textsubscript{j}})^2}{a^2}  + \frac{(V_{\perp\textsubscript{i}} - V_{\perp\textsubscript{j}})^2}{b^2} \leq 1,
\end{equation}
where the $r\textsubscript{lim}, a, b$ are free parameters, making an elliptical selection in V-plane owing to different intrinsic dispersions of $\Delta V_{\parallel}$ and $ V_{\perp}$ for the clusters under study. Stars whose entire neighbourhood lay inside the five-dimensional phase-space defined by Eq.~\ref{eq: space} and Eq.~\ref{eq: vel} were selected. This effectively dropped the sources close to the phase-space boundaries. This gave us the basic sample (BS) for every cluster.

The $k$-neighbourhood for every star is the number of five-dimensional neighbours that star $i$ has in this phase space. Initially, all the sources with $k$ $\leq$ 5 for the clusters within 350 pc and $k$ $\leq$ 4 for the clusters farther than 350 pc were considered to be local Galactic background sources that contribute to noise. For these sources, the mean value ($\lambda$) of five-dimensional neighbours was calculated, and the Galactic background was modelled as a Poisson distribution with this $\lambda$. 
This particular step was a modification to the approach of \citetalias{2019A&A...627A...4Roser_Praesepe} to compensate for the phase-space volume we considered, which was larger than the selection made by \citetalias{2019A&A...627A...4Roser_Praesepe} for the Praesepe. We subsequently found a greater number of co-phase-space overdensities around the clusters we studied, which effectively increased the number of stars with $k$ > 5 in the BS. These co-phase-space overdensities were the other clusters that lay close-by to the cluster we studied.
In this case, the mean number of neighbours for all the stars in the BS was quite high for every cluster, and the Poisson distribution with this mean overestimated the local Galactic background noise. Our approach yields a better estimate for such cases. We then calculated the Poisson distribution probability mass function $P\textsubscript{$\lambda$}(k)$, $N\textsubscript{exp}(k)$, $N\textsubscript{obs}(k)$, $p\textsubscript{cont}$ from the same definitions as \citetalias{2019A&A...627A...4Roser_Praesepe}.  Finally, we only considered the sources with $k$-neighbours for which $p_{\rm{cont}}$ was nearly 0.10 ($\equiv$10 \%) or slightly larger. For all the clusters, this occurred for stars with $k =$ 6. Some manual cuts were performed to remove co-phase-space overdensities due to other clusters that left the cluster and the tidal tails, if they existed, in the phase space. The $p_{\rm{cont,max}}$ of the cluster and tidal tails owing to $k =$ 6 are also reported in Table~\ref{Table: Tidal tails}.

The selection of free parameters was a crucial step in this analysis. It was made with a trial-and-error method with different permutations. 
The combination that recovered a maximum signal with a minimum background noise and eventually traced the tails up to a good extent without being affected by field stars, was chosen to be the best \footnote{Stars with $k$ = 0 and $k$ = 1 should have $p_{cont}$ close to one when the neighbour-finding parameters are optimum. Larger parameters than those would cause $p_{cont}$ for $k$ = 0 or 1 to deviate significantly from one. In this case, the sample is unreliable as the background noise for all values of $k$ is enhanced and may be detected as a false signal, with $k$ greater than or equal to 6.}.
The final free parameters for all clusters are listed in Table~\ref{Table: Tidal tails}. The values of $a$, $b$, and $r\textsubscript{lim}$ do not only depend on the intrinsic dispersions of cluster members, but also varied greatly with ambient phase-space density of the field stars around the clusters. We therefore considered a $k$ value of 4 for the neighbours of Galactic background sources beyond 350 pc. Some cases of interference, especially in velocity-space due to another cluster, were also seen. We were unable to map out the tails reliably for these clusters (e.g. OC CWNU 1084).
Closer to the Galactic disk, the background density introduces noise, and a smaller $r\textsubscript{lim}$ was chosen to mitigate this. This optimal trade-off gives a sample of tails that is more reliable because the noise is lower, but it truncates tails to shorter distances from the cluster.
%  Similar kind of interference but due to higher background noise arises
%  in phase-space for the clusters closer to Galactic disk in which a
% compromise at smaller values of rlim is necessarily made affect-
% ing the overall extent of tidal tails recovered

\begin{comment}

To test the reliability of stars in the tidal tails found by the CP method, we employed another criteria by finding the distribution of stars in a phase space created by two orbital parameters: $Z-$component of angular momentum of a star Vs its total orbital Energy both of which are conserved quantities in a typical Galactic potential,  ($L\textsubscript{Z}-E$) phase space. The use of overdensities in the ($L\textsubscript{Z}-E$) space was demonstrated by \citet{Bonaca_2021}, \citet{2024ApJ...964..104Malhan} and many others to find the older population of stellar streams and proto-galactic filaments in the inner Galactic plane. This filtering works best for the populations of stars which follow a circular orbit around the galaxy. Though not all OCs follow a circular orbit and the ($L\textsubscript{Z}-E$) calculations are limited for the brightest stars in the Gaia data for which radial velocities are available, trend in the ($L\textsubscript{Z}-E$) space was set with these stars and marked the once significantly deviating as outliers. 
\vik{WE NEED TO DISCUSS THIS, HOW EXACTLY IT IS IMPLEMENTED}
\end{comment}

\subsection{SCCP method}\label{sec: SCCP}

The use of a model-dependent CCP method for tracing the tidal tails of the Hyades and of NGC 752 was pioneered and demonstrated successfully by \citet{2021A&A...647A.137Jerabkova} and \citet{2022MNRAS.514.3579Boffin}, respectively. Similar to these works, \citet{2024arXiv240618767Kos} developed probabilistic N-body models for more than 450 open clusters and traced their tidal tails. As mentioned by \citet{2022MNRAS.514.3579Boffin}, the CCP or any N-body model-dependent works assume that observations follow the model(s), but this is not necessarily true. We developed a model-independent method called SCCP as an alternative to the model-dependent CCP.

\citet{2021A&A...647A.137Jerabkova} showed that the distance-dependent scatter in $V_{\parallel}$ for models of a Hyades-like cluster can span up to 40 km s$^{-1}$, with stars farther away from the cluster centre moving relatively faster. This trend in distance from the cluster centre, $R - R\textsubscript{cl}$ and $V_{\parallel}$, can be expressed with a linear function $\Lambda(R - R\textsubscript{cl})$. Correcting for $V_{\parallel}$ with $\Lambda(R - R\textsubscript{cl})$ by adopting $V_{\parallel \Lambda}$ = $V_{\parallel}$ - $\Lambda(R - R\textsubscript{cl})$ as a phase-space coordinate instead of the classical $V_{\parallel}$ yielded a distribution of stars that was much more compact in velocity space, and it performed better in finding the phase-space neighbours. This proved advantageous for finding the tails up to larger extents. 

For OCs Theia 517, Ruprecht 147, Stock 10, Platais 10, Collinder 350, Alessi 3, and Melotte 25, we observed a similar trend for the tails traced by the CP method, where the dispersion in $V_{\parallel}$ was linearly related to $R- R\textsubscript{cl}$. A linear fit in $V_{\parallel}$ Vs $R- R\textsubscript{cl}$ was performed for the CP-traced tails of these clusters to determine the correction factor $\Lambda\textsubscript{Self}(R - R\textsubscript{cl})$. The same neighbour-finding parameters $a, b, and r_{\rm{lim}}$ as the optimum ones found for the CP were used for the SCCP.    
% Adopting the corrected velocity similar to \citet{2021A&A...647A.137Jerabkova} for neighbour finding gave us the advantage to search for the tails to larger extent $(R -R\textsubscript{cl})$ with greater number of stars than classical CP without any model-dependence. This is new SCCP method we propose. 

The novel SCCP method also has some caveats that we note here. Not all the clusters in our sample showed a linear relation in $V_{\parallel}$ versus $R- R\textsubscript{cl}$, and hence, this method is not applicable for every cluster. For a few OCs that fit within this constraint, the fit was not always perfect compared to the fit on the Hyades models in \citet{2021A&A...647A.137Jerabkova}. This was expected as the observed data have uncertainties and depend on the CP selection. Hence, even if the $\Lambda\textsubscript{Self}(R - R\textsubscript{cl})$ is reliable, the SCCP missed a few stars that were detected by the CP method. These SCCP-missed stars are dispersed away from the velocity space overdensity near (0,0) in the course of the $\Lambda\textsubscript{Self}(R - R\textsubscript{cl})$ correction. We therefore created a merged catalogue by combining sources that passed the CP and SCCP methods, and they are listed in Table~\ref{Table: Tidal tails}.
% while reporting statistics for these OCs in table XXX and CDS catalogue, joint CP + SCCP samples have been presented.

\subsection{$L_{Z}-E$ secondary phase-space test}\label{sec: LzE_test}

We also employed a test to confirm the reliability of the tidal tails found by using the CP or SCCP methods by calculating the parameters dependent on the orbits of the stars in the cluster and the tails. A secondary phase space of the $Z-$component of the angular momentum of a star and its total orbital energy ($L\textsubscript{Z}-E$) space was adopted, and the distribution of stars in the cluster and the two tails was compared. $L_z$ and E were calculated using the \citet{McMillan2017MNRAS.465...76M} Galactic potential and \textsc{galpy} \citep{Bovy2015ApJS..216...29B}.  $L\textsubscript{Z}$ and $E$ are conserved quantities in a typical axisymmetric galactic potential, and the use of overdensities in the ($L\textsubscript{Z}-E$) space was demonstrated by \citet{Bonaca_2021}, \citet{2024ApJ...964..104Malhan} and many others to determine the older population of stellar streams and proto-galactic filaments in the inner Galactic plane. The calculations of the secondary phase-space coordinates $L\textsubscript{Z}$ and $E$ were limited for the brightest stars in the Gaia data for which radial velocities are available. The results from this test are discussed in Section~\ref{sec: LzE_results}.

\begin{comment}
Phase-space boundaries of clone Praesepe clusters were defined in slightly different way, owing to its prior distribution. As these clusters had tails along $Y$- axis, we did not constrained $Y$-coordinate of phase-space. $X$-coordinate of was constrained to $X_\textsubscript{min,clone}$ $<$ $X$ $<$ $X_\textsubscript{min,clone}$. This would mean that the stars at $X$- boundaries of phase space would not have their entire neighbourhood included in the phase-space, but these stars were anyways so far away from the cluster and its tidal tails that they appear completely disconnected from cluster and its tails and would not be qualified as members of tails even if they have $k >$ 6. These cuts reduce the clone sample size to be analysed without affecting the overall extent of the tails.  
    
\end{comment}

\section{Results} \label{sec:results}

\begin{figure}[h]
    
    \includegraphics[width=1\linewidth, height =1\linewidth]{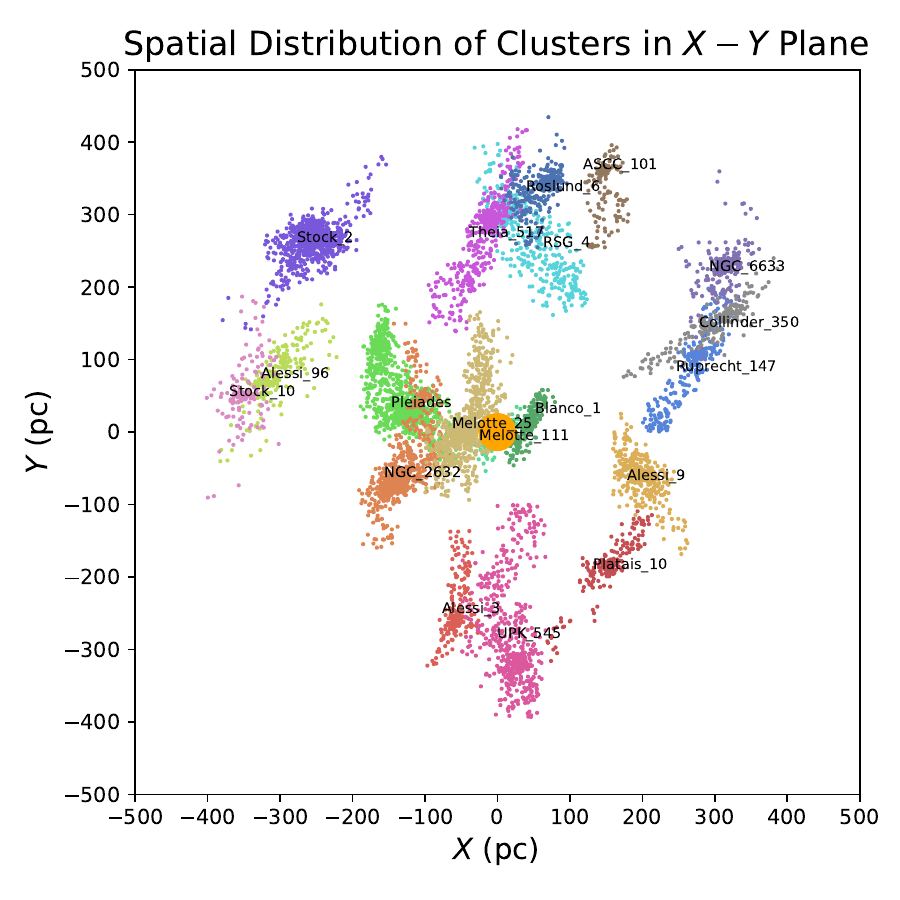}
    \caption{$X-Y$ plot of all the clusters and their tidal tails. The Galactic centre is in the positive $X$ direction (to the right), the Galactic rotation is towards the positive $Y$ direction (upwards), and the Sun is at the origin. A 3D interactive display can be found in the supplementary material of this article.}
    \label{fig:All_in_one_XY}
\end{figure}

In total, we analysed 21 OCs, 19 of which show clear evidence for the existence of tidal tails, and 5 of which are discovered for the first time. The comparison with previously known clusters shows that the tails we found extend farther and have a significantly larger number of stars. Figs.~\ref{fig:ASCC_101} to~\ref{fig:UPK_545} document the results for individual clusters, and Figure~\ref{fig:All_in_one_XY} shows the $X-Y$ distribution for all clusters. Fig.~\href{https://doi.org/10.5281/zenodo.14651597}{Z20} shows the same in Galactic coordinates. The numbers we report in Table~\ref{Table: Tidal tails} were subjected to specific data-quality filters (see section~\ref{sec:data}) and the method (see section~\ref{sec:methods}).

We also confirmed the existence of the halo surrounding NGC 6475, which was discovered by \citet{2021ApJ...912..162Pang}. Although an extended structure around CWNU 1084 was found, we were unable to confirm its association with the cluster as its phase space is most strongly affected by interference from nearby clusters. A catalogue of the cluster and tidal tail members is available at the CDS. 

\begin{comment}

\begin{figure*}
    \includegraphics[width=1\linewidth, height = 1\linewidth]{figures/All_in_one_XY.pdf}
    \caption{$X-Y$ plot of all the clusters and their tidal tails analysed in this work. The Galactic centre is in the positive $X$ direction (to the right), the Galactic rotation is towards the positive $Y$ direction (upwards) and the Sun is at the origin. A 3D interactive display of the same can be found in the supplementary material of this article.}
    \label{fig:All_in_one_XY}
\end{figure*}
\end{comment}

\subsection{The $L_Z-E$ phase-space test. \label{sec: LzE_results}}

The distribution of a cluster and its tidal tails in the $L_{Z}-E$ plane are expected to follow a narrow arc-shaped structure based on the cluster orbit around the Galaxy (e.g. panel (i) of Figs.~\ref{fig:ASCC_101} to~\ref{fig:UPK_545}). 
We found a handful of outliers that deviated from the mean distribution in all clusters. These co-moving and co-spatial contaminants were located in and outside the tidal radius. They were not removed from the final catalogue, however, because there are only a few of them and we were unable to perform similar cuts for stars without a radial velocity.
The comparison between the distribution of stars within the cluster and the tidal tails did not show significant differences, and the arc-shaped structures of these three regions match well. This supports the identification of tail members.

%The distribution of a cluster and its tidal tails in the $L\textsubscript{Z}-E$ should follow a very narrow, arc-like structure reflecting the deviation from perfectly circular orbit, without owing to scatter in the velocities. The stars ejected out from the cluster may possibly have orbits around the GC different than the cluster and comoving tails in whole. For almost every cluster in our selected sample we found such outliers far deviating from the arc-like structure. The comparison between the distribution of stars within the cluster (inside the tidal radius) with the once in the tails did not show significant differences and arc-like structures of these three regions match reliably to a certain degree. This test confirmed indeed that the correlated structures in the primary five-dimensional phase space were clusters and their respective tidal tails.

\subsection{Improvement through the of CP + SCCP combined sample}
For the clusters and their tidal tails for which we were able to use the SCCP method described in Section~\ref{sec: SCCP}, we formed combined samples of stars traced by either CP or SCCP. These samples comparatively performed better overall than individual samples in terms of the number of stars and the extent of the tidal tails. Moreover, there was no significant change in the Poisson statistics in the SCCP basic sample when $V_{\parallel \Lambda}$ was adopted instead of $V_\parallel$, and $p_{\rm{cont}}$ for any given $k$ remained nearly the same. This shows that the combined samples are as reliable as the CP samples with similar contamination.

\begin{figure*}
    
    \includegraphics[width=1\linewidth, height =1\linewidth]{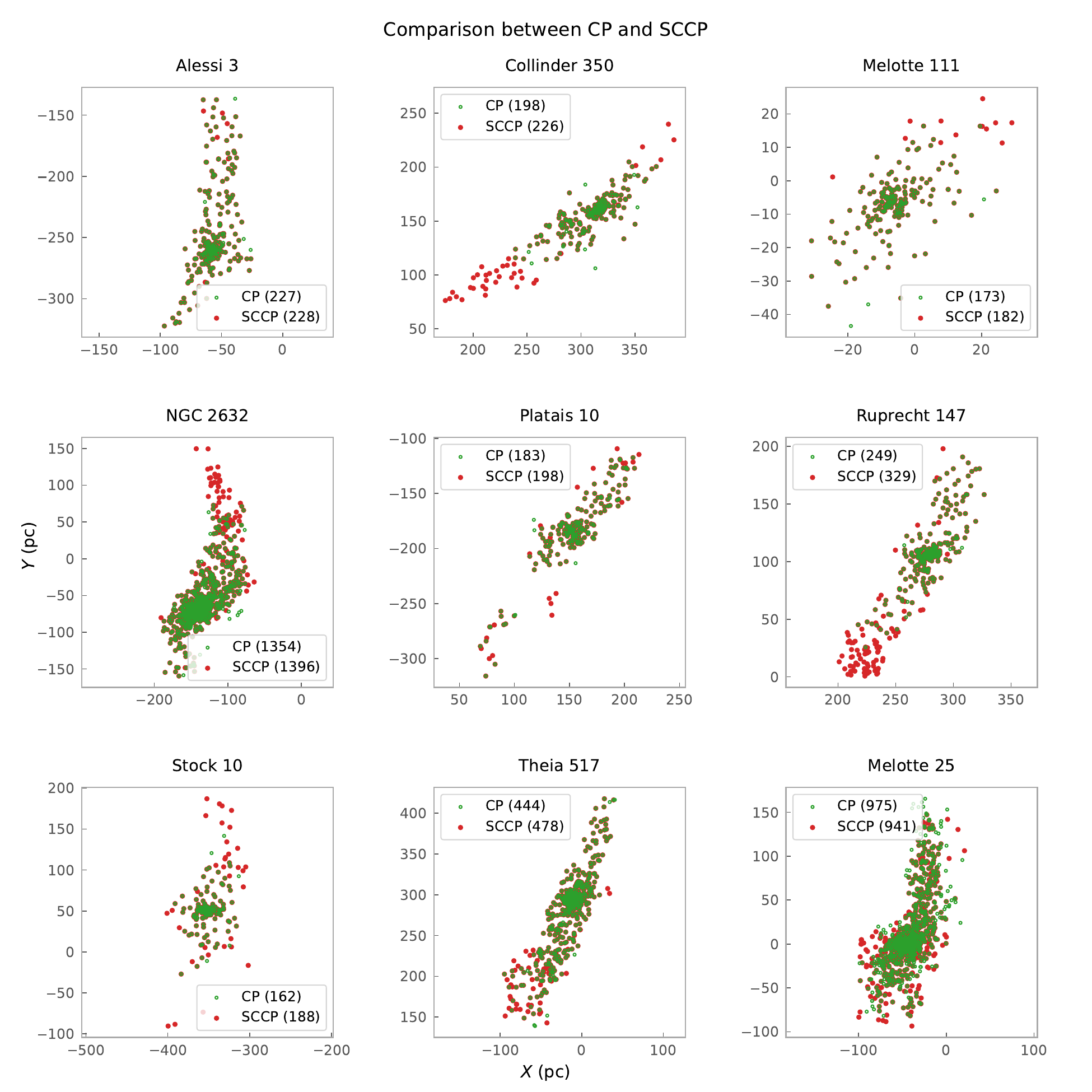}
    \caption{Plots in $X-Y$ coordinates for the comparison between the tidal tails traced by the CP and SCCP methods. }
    \label{fig: CP_SCCP_Comp}
\end{figure*}

The percentage increase in tracing the number for stars for the combined sample compared to the CP sample varied from 2\% to 34\% in our clusters. For Ruprecht 147, the increase is highest at 34.5\%, with the recovery of a 50 pc longer trailing tail. For Collinder 350 and Stock 10, an improvement by 18.7\% and 20.7\% was observed, respectively, with greater recoveries and extents of both tails. For NGC 2632, the combined sample had a 100 pc longer leading tail, with a percentage increase in the number of stars of 5.7\%. For Alessi 3 (2.6 \%), Melotte 111 (6.9 \%), Platais 10 (11.5 \%), and Theia 517 (10.1 \%), the recoveries were greater within the same extent of both tails. These results are summarised in Figure~\ref{fig: CP_SCCP_Comp}.

\subsection{Corrected samples}\label{sec: Corrected Samples}

The overall statistics of the recovery of the tidal tails found in this work is summarised in Table~\ref{Table: Tidal tails}. As the tidal tails are extended systems that in some cases span up to more than 100 pc, and owing to their stellar distribution in space with respect to the Sun, the samples are always biased due to the distance-dependent sensitivity in the Gaia $G$ magnitude. In most cases, a larger number of stars are therefore detected in the tail nearer to the Sun than farther away. It is impossible to eliminate this bias completely, but it can be minimised by assuming a cut-off $M_G$ magnitude and considering the stars with $M_G$ below this cutoff. We calculated the cutoff as
\begin{equation} \label{eq:MGlim}
\begin{split}
    M_{\text{G, Lim}} = \min\big( &\max(M_{\text{G, Lim, tidal}}), 
                                  \max(M_{\text{G, Lim, leading}}), \\
                                  &\max(M_{\text{G, Lim, trailing}}) \big) - 1
\end{split}.
\end{equation}

For a given cluster, we found the stars within both tails up to different distances. We truncated the longer tail up to the distance of the shorter tail from the cluster and only considered the stars that passed $M_{\text{G, Lim}}$. We refer to these as the corrected samples and report their statistics in the same Table~\ref{Table: Tidal tails}. Figures of individual clusters and their tails can be the appendix~\ref{sec:Fig_Desc} Figs.~\ref{fig:ASCC_101} to~\ref{fig:UPK_545}. The comments on the individual clusters and the comparison with literature on the basis of the overall extent, that is, the full tip-to-tip length and the number of stars in the cluster and the tails, is given below. Figure~\ref{fig:Comp_with_Lit} shows the comparison of the tidal tails of different OCs found in this work with previous studies. We recall that the more distant tail has poorer trigonometric data. We will return to their additional bias in a forthcoming contribution (Risbud et al., \textit{in prep.})

\begin{comment}
    
\vik{Out of the 22 clusters, we detected tidal tails in XX clusters. YYY,YYY,YYY did not show evidence for tidal tails using the CP method due to ZZZZ.}

\vik{The overall statistics of recovery are given in Table~\ref{Table: Tidal tails}. The complete list of cluster and tail members is available at CDS.}

\vik{We can recover members till XXX pc, however the recovery rate drops for farther clusters (e,g YYY)}

\vik{After correcting for magnitude, the ratios change (or don't)}
\end{comment}

\subsection{ASCC 101}
The sample of the cluster and its tidal tails from \citet{10.1093/mnras/stac2906} consists of 103 stars, very few of which are outside the tidal radius. We found a sample for the same with 144 stars in total, with the trailing tail extending to 100 pc from the cluster. In the tidal radius and near the tidal region, 58 stars cross-match from both samples. Due to the position of the cluster at $(X,Y,Z) = (145, 360, 80)$ pc, we were unable to find many stars in the leading tail, which extends up to 35-40 pc from the cluster away from the Sun.

\subsection{Alessi 3}
The tidal tails of Alessi 3 were mapped by \citet{2022ApJ...931..156Pang}, who found a total of 165 member stars in the cluster and tails. The extent of the leading and trailing tail was less than 30 pc from the cluster centre. We found a total of 233 members, 124 (74.5 \%) of which cross-matched within an extent of 30 pc. Our work revealed a much larger extent of the tidal tails, with leading and trailing tails extending up to 120 pc and 60 pc from the cluster centre, respectively.

\subsection{Blanco 1 (zeta Scl)}
\citet{2020ApJ...889...99Zhang} mapped the tidal tails of Blanco 1 with the data from Gaia DR2 using unsupervised machine-learning in five-dimensional phase space. They reported 644 members within the tidal radius and the tails. Similarly, \citet{2021A&A...645A..84Meingast} reported 494 members. We found 554 members with 517 (80 \%) and 424 (85\%) in common with the two studies, respectively. The overall extent of the tails found by all three is approximately 50-60 pc for each tail from the cluster centre. 

\subsection{Collinder 350}
Similar to Alessi 3, \citet{2022ApJ...931..156Pang} revealed 156 stars associated with Collinder 350. Its tidal tails extend to less than 30 pc from the cluster. We mapped the same tails up to an extent of 50 pc with a total of 235 stars, 106 (64.9 \%) of which were found to be in common with the \citet{2022ApJ...931..156Pang} survey in the 30 pc region around the cluster. 

\subsection{Melotte 111 (Coma Berenices, Collinder 256)}
The tidal tails of Melotte 111 were discovered by \citet{2019A&A...624L..11Furnkranz, 2019ApJ...877...12Tang,2022ApJ...931..156Pang}. \citet{2019A&A...624L..11Furnkranz} used the approach of overdensities in galactocentric velocity space to reveal the tidal tails. They found 214 stars with an overall extent of about 50 pc. A cross-match using the Gaia source IDs with our sample of 212 stars gave 170 (80.2 \%) stars in common in both works.

\citet{2019ApJ...877...12Tang} used a similar clustering algorithm in five-dimensional phase space to reveal an overdensity of the cluster and its tails. They found 197 stars within 50 pc, 164 (83.2 \%) of which match our sample. 

While searching for the overdensities, we also detected group X identified by \citet{2019ApJ...877...12Tang}. We further confirmed that this group X was identified by \citetalias{2024arXiv240305143H} as the moving group HSC 759. A comparison between the basic properties from \citetalias{2024arXiv240305143H} showed that HSC 759 is younger, with log\textsubscript{10}$(\frac{age}{yr})$ = 8.2. The mean differential space velocities of HSC 759 compared to Melotte 111 are (-8.007, 61.603, 76.324) km s$^{-1}$. Although the overdensities of Melotte 111 and HSC 759 appear to be related in phase space, HSC 759 is not associated with the leading tail of Melotte 111 because their ages and space motions are different. It is an example of phase-space interference. 

\subsection{Melotte 22 (Pleiades, Collinder 42, M 45, Theia 369)}

The tidal tails of Melotte 22 were extensively mapped out and studied in great detail by various methods in the literature by \citet {2021RNAAS...5..173Li,2021A&A...645A..84Meingast, 2021MNRAS.505.1607Bhattacharya}. A total of 1407, 1177, and 1410 stars are reported in the cluster and the tidal tails. The extent of the overall structure found by these three works is very different.  \citet{2021RNAAS...5..173Li} found stars in the tidal tails up to a few parsec outside the tidal radius, with an overall extent of less than 70 pc around the cluster with a more or less a circular distribution when projected onto the $X - Y$ plane.  \citet{2021MNRAS.505.1607Bhattacharya} found a 100 pc long structure, where the region inside the tidal radius has a circular projection, with the leading tail directed along the $Y -$ axis in the $X-Y$ plane and a slight tilt to the side of the Galactic centre. Conversely, \citet {2021A&A...645A..84Meingast} found a structure with a similar extent and a leading tail tilted more towards the Galactic anti-centre. We found 2029 stars associated with the Pleiades and their tails, with an overall tip-to-tip extent in the $X-Y$ plane of more than 200 pc and the leading tail tilted  towards the Galactic anti-centre. The Gaia source ID cross-matches with all three works gave 1240, 1117, and 1180 stars in common. We also found two moving groups within the tidal tails of the Pleiades. We discuss them in Section~\ref{sec: Moving Groups}.

From the findings of \citet{2023A&A...679A.105Vaher}\footnote{See Section 2.1 from \citet{2023A&A...679A.105Vaher} for the definitions of $p_f$ and $p_{min}$.}, we took the sample of the Pleiades with $p_f$ $\geq$ $p_{min}$ = 0.1. 
The comparison of our sample with the sample from \citet{2023A&A...679A.105Vaher} shows a striking resemblance. The leading tail is tilted towards the Galactic anti-centre and the trailing tail is tilted towards the Galactic centre. The extent of the trailing tail from both works is quite similar, but \citet{2023A&A...679A.105Vaher} found a shorter leading tail and did not detect the moving group we found.
Out of 462 stars from \citet{2023A&A...679A.105Vaher}, 360 stars (78\%) cross-match with our sample.

\subsection{Melotte 25 (Hyades)} 
\citet{2019A&A...621L...2Roser_Hyades} discovered the tidal tails of the Hyades with a total of 972 stars in the tails and the cluster. Our CP sample consists of 975 stars, which agrees very well with the literature. We considered a larger phase space in $Z$ coordinate, 35 pc in both directions around the Hyades, compared to 20 pc taken by \citet{2019A&A...621L...2Roser_Hyades}. The CP + SCCP combined sample includes 1069 stars, 753 of which were found to be in common in both works. The extent from the cluster of the leading and trailing tail recovered in the two works was found to be the same, 170 pc and 70 pc.

\begin{figure}[h]
    
    \includegraphics[width=1\linewidth, height =1\linewidth]{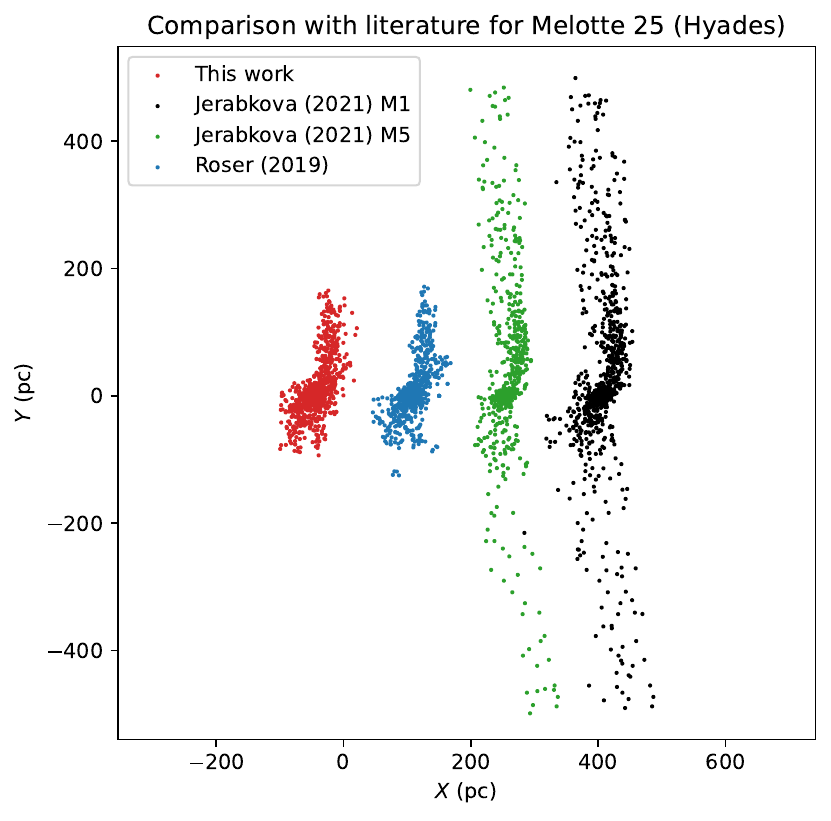}
    \caption{Plot in $X-Y$ coordinates for a comparison of Melotte 25 (Hyades) and their tidal tails with the literature. The coordinate system is the same as in Fig. \ref{fig:All_in_one_XY}. The studies from the literature are plotted with offsets of 150, 300, and 450 pc in the $X$ coordinate for clarity. }
    \label{fig:Melotte_25_comparison}
\end{figure}

\citet{2021A&A...647A.137Jerabkova} found 1109 and 862 stars in the Hyades and its tidal tails with the CCP method from models M1 and M5, respectively. Our sample only extends to 170 pc in the leading and 70 pc in the trailing tail direction, compared to their sample, which reaches 400 pc in both directions. In the overlap between the two, we found 703 and 615 stars in common in both works for models M1 and M5, respectively. The comparison of our sample of Melotte 25 with the literature is shown in Fig.~\ref{fig:Melotte_25_comparison}.

It has been a long-standing problem that there are fewer white dwarfs than expected in the Hyades (and other galactic OCs). This is called the white dwarf deficit (see e.g. \citealt{1977A&A....59..411Weidemann, 1992AJ....104.1876Weidemann, 2001AJ....122..257Kalirai}). Preferential ejections of white dwarfs in the course of dynamical processes were discussed as a possible solution to this problem (see e.g. \citealt{2003ApJ...595L..53Fellhauer} and reference therein). We observed a clear sequence of 19 probable white dwarf candidates (see panel (d) of Fig. \href{https://doi.org/10.5281/zenodo.14651597}{Z8}  ), 13 of which are located in the tails. 

\subsection{NGC 2632 (Praesepe, M 44, Melotte 88) }
The tidal tails of Praesepe were traced out by \citetalias{2019A&A...627A...4Roser_Praesepe} using Gaia DR2 with the same CP method with which \citetalias{2019A&A...627A...4Roser_Praesepe} found that 1357 stars belong to the cluster and the tails, with a contamination probability $\leq$ 0.561 or $k$ $\geq$ 3. With our slight modification to the method from \citetalias{2019A&A...627A...4Roser_Praesepe} (see the Section~\ref{sec:finding_overdensity}), we found a total of 1428 stars with $k$ $\geq$ 6, 1208 (89.0 \%) of which match \citetalias{2019A&A...627A...4Roser_Praesepe}. 
Moreover,  1354 stars in our study were traced by the CP method, which agrees very well with \citetalias{2019A&A...627A...4Roser_Praesepe}, with 1181 (87.2 \%) cross-matches.
The extent of the trailing tail found by us is somewhat shorter (approximately 100 pc) than was found by \citetalias{2019A&A...627A...4Roser_Praesepe}. 
We also confirm the existence of the separately identified moving group HSC 1244 within the leading tail of NGC 2632, which was found by  \citetalias{2019A&A...627A...4Roser_Praesepe}. This is discussed in Section~\ref{sec: Moving Groups}.

\subsection{NGC 6633 (Collinder 380, Melotte 201)}
\citet{2022ApJ...931..156Pang} analysed the three-dimensional morphology of NGC 6633, but failed to find any evidence for its tidal tails. We identified the extended structure and tidal tails of NGC 6633 with a total of 358 stars. Due to its larger distance of approximately 400 pc from the Sun, we were only able to recover a few stars in its tidal tails.

\subsection{Roslund 6 (RSG 6)}
\citet{10.1093/mnras/stac2906} attempted to find the tidal tails of Roslund 6, where 230 member stars were identified in total. The overall extent of the tails was only a few parsec outside the tidal radius, with very few stars. We mapped the stars up to a distance of 100 pc from the cluster in the leading and trailing tails, with a total count of 357 stars. Inside the tidal radius and near the tidal region, 164 stars were found by cross-matching with \citet{10.1093/mnras/stac2906}.

\subsection{Theia 517 (Collinder 438, M 39, Melotte 236, NGC 7092)} 
The tidal tails of Theia 517 were studied in detail by \citet{2021A&A...645A..84Meingast} and \citet{10.1093/mnras/stac2906}. \citet{2021A&A...645A..84Meingast} found a total of 351 stars, and the extent of the leading and trailing tail was about 100 and 80 pc from the cluster centre, and 308 stars were found by \citet{10.1093/mnras/stac2906}, with both tails extending for a few parsec outside the tidal radius. Our sample has 489 stars, which is significantly larger than any of the literature studies. The extent of the leading and trailing tail we mapped is larger by 20 pc and 70 pc than \citet{2021A&A...645A..84Meingast}. A cross-match using the Gaia source IDs of our sample with \citet{10.1093/mnras/stac2906} and \citet{2021A&A...645A..84Meingast} gives 86 and 267 (76.0\%) stars in common, respectively.

\subsection{Halo around OC NGC 6475 }
We confirm the existence of a halo of stars around OC 6475, which was discovered by \citet{2022ApJ...931..156Pang}. We found 1005 stars in the cluster and halo, 866 of which cross-match with 1157 stars found by \citet{2022ApJ...931..156Pang}.

\begin{figure*}
    \includegraphics[width=1\linewidth, height =1\linewidth]{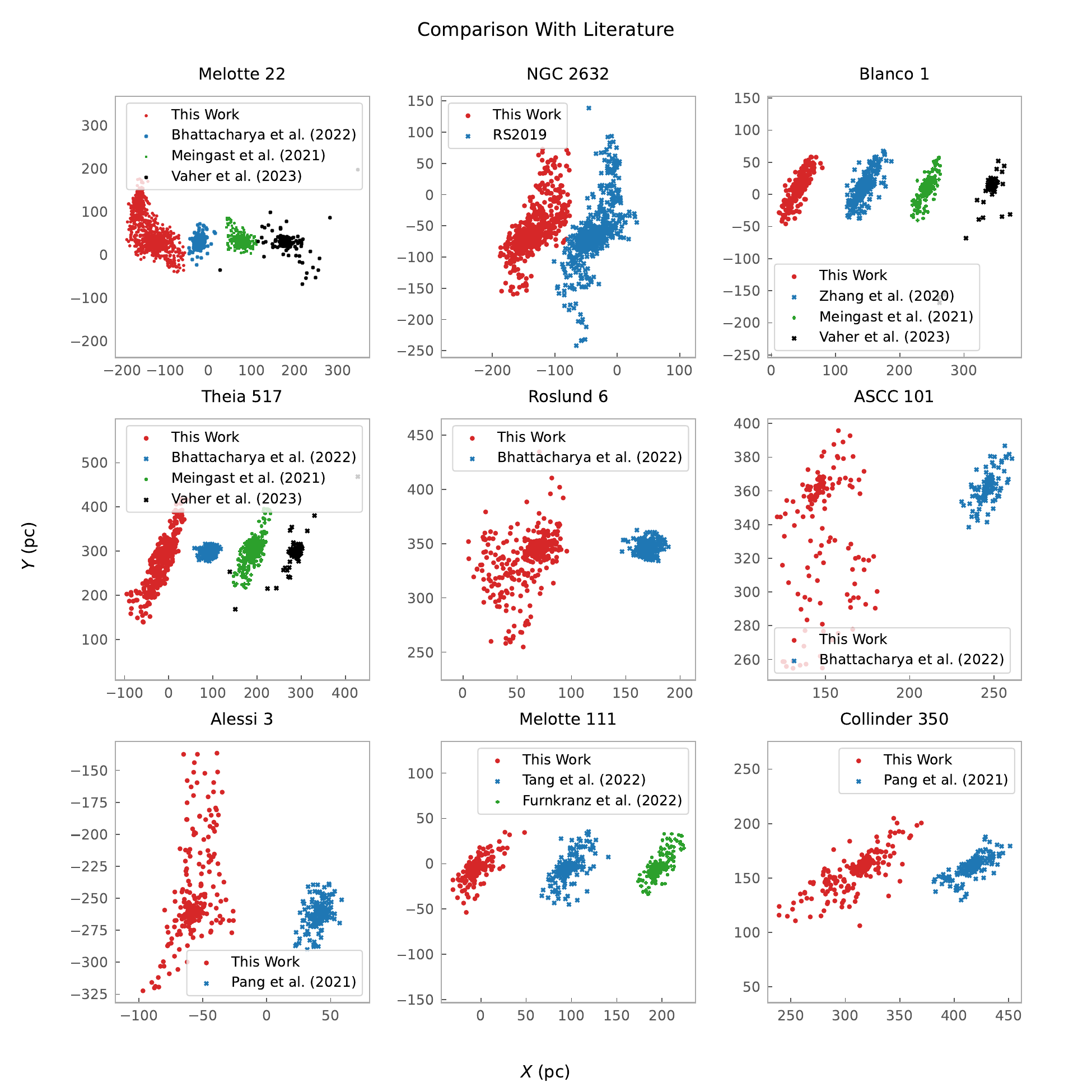}
    \caption{Plots in $X-Y$ coordinates for the comparison of different clusters and their tidal tails with the literature. The coordinate system is the same as in Fig. \ref{fig:All_in_one_XY}. Successive +100 pc offsets were added to studies from the literature for clarity.}
    \label{fig:Comp_with_Lit}
\end{figure*}

\section{Discussion} \label{sec:discussion}

\subsection{Analysis in $\phi_1$ - $\phi_2$ space}\label{sec: phi1-phi2}

The projections of spatially three-dimensional tidal tails in the equatorial system ($\alpha$ - $\delta$) appear to be elongated for a few dozen degrees on the sky.
The usual equatorial coordinates can be spherically rotated into a new system that is aligned with this elongation along the tail. We call this new system ($\phi_1$, $\phi_2$). The goal of this transformation is to align the latitude $\phi_1$ along the elongated tail and the longitude $\phi_2$ to minimise the dispersion, which remains perpendicular to $\phi_1$. This helps us to analyse the kinematical and astrophysical parameters of the tails as a function of $\phi_1$. Similar elongation-aligned coordinates were used previously by \citet{2010ApJ...712..260Koposov}, \citet{2019MNRAS.485.4726Koposov}, \citet{2019MNRAS.487.2685Erkal},
\citet{2023MNRAS.521.4936Koposov} and many others to analyse Galactic stellar streams (where they are called stream-aligned coordinates), which are several dozen kiloparsec long and distant.

We used the \textsc{gala} \citep{2017JOSS....2..388Price} Python package for the spherical rotation from ($\alpha$, $\delta$) to ($\phi_1$, $\phi_2$). It needs a great circle with the pole ($\alpha_{\rm{pole}}$, $\delta_{\rm{pole}}$) to be defined in ($\alpha$, $\delta$). We estimated this by lying a grid of tests ($\alpha_{pole}$, $\delta_{pole}$) on the sky with a resolution of 1\arcdeg and minimising the squared distance from the $\phi_2$ = 0 great circle. The poles determined for all the clusters and their tidal tails are given in Table \ref{Table: phi_org_pole}. The origin
\footnote{\texttt{gala.coordinates.GreatCircleICRSFrame.from\_pole\_ra0} was used for the transformations with argument \texttt{ra0} = $\alpha_{pole}$, except for Blanco 1, where it was given to be 0\arcdeg due to its position on sky. The argument \texttt{origin\_disambiguate} was always passed to be ($\alpha_{pole}$,$\delta_{pole}$).}  
of the ($\phi_1$, $\phi_2$) system was chosen to be the mean ($\alpha$, $\delta$) of the stars inside the tidal radius of the cluster.

The proper motions ($\mu_{\alpha} \rm{cos} \delta $, $\mu_{\delta}$) were initially corrected for the solar reflex motion with \texttt{gala.coordinates.reflex\_correct}\footnote{\texttt{gala.coordinates.reflex\_correct} needs the radial velocities of stars as well. For all the stars independently of the radial velocities being available or not, 0 km s$^{-1}$ was passed to avoid the differential corrections.}.
These solar reflex-corrected ($\mu_{\alpha, refxcorr} \rm{cos} \delta $, $\mu_{\delta, refxcorr}$) were transformed into ($\mu_{\phi_{1}} \rm{cos} \phi_2 $, $\mu_{\phi_{2}}$), which are along the new tail-aligned system.

In the $\phi_1$ versus $\phi_2$ plots (subplot (m) of Figs.~\ref{fig:ASCC_101} to~\ref{fig:UPK_545}), the tails of clusters near $Y$ $\approx$ 0 appear to a have constant spread in $\phi_2$ throughout the elongation $\phi_1$. For clusters at higher |$Y$|, for which the tails in $X-Y$ are along the line of sight, the ends closer to $Y$ = 0 appear to be more dispersed in the  $\phi_1$ versus $\phi_2$ plots. This is a pure projection effect. 

The analysis in $\phi_1$ versus $\mu_{\phi_{1}} \rm{cos} \phi_2 $ (panel (j) of Figs.~\ref{fig:ASCC_101} to~\ref{fig:UPK_545}) shows that in the increasing direction of $\phi_1$, for the tails in $Y$ < 0,  $\mu_{\phi_{1}} cos \phi_2 $ monotonically increases, and for the tails in $Y$ > 0, it monotonically decreases. The few exceptions to these are Alessi 96, Blanco 1, Melotte 111, and Stock 10, where no trend was seen. Interestingly, for Melotte 25 and NGC 2632, the parts of their tails with $Y$ > 0 monotonically decreases, and the rest with $Y$ < 0 monotonically increases, forming an arc-like shape in the $\phi_1$ versus $\mu_{\phi_{1}} cos \phi_2 $ plot. The bulk proper motions of the tails and the cluster always point in the direction of the leading tail.

The tail-aligned proper motions of stars $\sqrt{(\mu_{\phi_{1}} cos \phi_2)^2+ (\mu_{\phi_{2}})^2 }$ are expected to be along the elongation in $\phi_1$, which was found to be the case for most of the clusters. For a few exceptions such as Blanco 1, Collinder 350, Melotte 111, and RSG 4, however, the tail-aligned stellar proper motions deviate strongly from the tail elongation in the $\phi_1$ direction. This is inherited from the deviation of their Galactocentric velocities from the elongation along $Y$ in three dimensions.  

\begin{comment}
As shown by \citet{2023MNRAS.521.4936Koposov}, the Large Magellanic Cloud attracts a part of Orphan stellar stream closer to it. Hence the proper motions in the stream-aligned coordinates for that part deviate slightly from the stream orientation and point towards the Large Magellanic Cloud. We suspect a similar reason for the exceptions for which another closer and relatively massive cluster may deviate the ($\mu_{\phi_{1}} cos \phi_2 $, $\mu_{\phi_{2}}$) of the cluster and its tails towards it. 
\end{comment}

Although radial velocities (RVs) were only available for the brighter stars in the clusters and tails, we analysed them as a function of $\phi_1$, as shown in panel (k) of Figs.~\ref{fig:ASCC_101} to~\ref{fig:UPK_545}. Stars with very high |RVs| inside or outside the tidal radius are most likely outliers in the sample. A slightly tilted sequence is formed by the stars predominantly in the tails, mostly with individual stellar RVs close to the mean RV of the cluster. In most of the OCs, stars in their respective trailing tails have higher RVs than those in the leading tails. If there is no significant detection bias in the data, this physical effect may probe the gradient of the local Galactic potential around the OCs, where escape from the cluster is comparatively harder on the side of the trailing tail, and only the stars with higher velocity are therefore seen there.

The OCs and their tails that lie comparatively closer to the Sun, such as NGC 2632 (Praesepe), Melotte 22 (Pleiades), and Blanco 1, have better statistics and a greater number of stars with RVs, where tilted sequences are distinctly seen. In others, such as Theia 517, RSG 4, and Ruprecht 147, this is less distinct, but still clearly visible.

\subsection{Possible biases and incompleteness}\label{sec: Possible biases }

The N-body computer simulations of tidal tails in general clearly show the extent of the tidal tails up to kiloparsec scales for older clusters such as the Hyades \citep[e.g.][]{2020A&A...640A..84Dinnbier,2021A&A...647A.137Jerabkova,2023A&A...671A..88Pflamm-Altenburg}. Younger OCs are expected to have extents shorter than one kiloparsec. The CP method is limited in this aspect because the tail motion deviates from the cluster at larger distances and there are biases in tracing the tidal tails. It yields the best results for massive nearby OCs that trace their tails to an extent of 100-150 pc (e.g. Melotte 22 in Fig.\href{https://doi.org/10.5281/zenodo.14651597}{Z7}), which are significantly overdense above the background in their respective five-dimensional neighbourhoods. The less massive OCs intrinsically have relatively fewer stars in the tidal tails, and their density decreases along the tails. It was not possible to trace them equally well.

In addition to the population, the distance of the cluster and its tails has the largest effect on the recoveries. The spatial volume density of the previous cluster member stars is a decreasing function of the distance from the cluster. The field density for five-dimensional neighbourhoods of distant OCs is intrinsically lower than that of the nearer OCs. This reduces the number of stars with higher $k$ for these. This is reflected in the recoveries also for OCs within distances $\approx$ 400 pc. For example, Collinder 350, NGC 6633, Stock 10, and Stock 2 evidently have poor recoveries in the tidal tails. As the tidal tails are generally along the galactocentric Y-axis, the clusters with $l$ $\approx$
90 or 270 have one tail closer to the Sun than the other. Hence, these clusters are likely significant biased due to differential recoveries in the nearer and farther tail. This is evident in Figure~\ref{fig:All_in_one_XY}, where the nearer trailing tails of ASCC 101, RSG 4, and Roslund 6 are better traced than their respective leading tails. Similarly, the leading tails of Alessi 3 and UPK 545 are recovered comparatively better than the farther trailing tails. More analysis is required with simulated data to understand these biases in detail. %, article regarding which is in preparation. (Risbud et al., \textit{in prep.})

\subsection{Moving groups within the tidal tails}\label{sec: Moving Groups}

In several cases, two or more separated but neighbouring phase-space overdensities are detected significantly above the background noise of field stars. In these cases, it is critical to distinguish whether they are associated with each other. The selection of the single strongest overdensity of the cluster affects the overall recovery and extent of the tidal tails. In contrast, a faulty association of two or more overdensities increases the contamination probability without being quantified in $p\textsubscript{cont}$. We verified these associations by similarities in the galactocentric positions, galactocentric velocities, CP velocities, ages, and CMDs. The minor overdensities at distances of a few dozen parsec from a cluster in the spatial coordinates were catalogued in \citetalias{2024arXiv240305143H} as separate moving groups. We found the missing links by associating them with clusters in five-dimensional phase space, where the moving groups were a part of the tidal tails of Melotte 22 and NGC 2632. These results are summarised in Table~\ref{Table: Moving_Groups}. In contrast, we found some clusters whose tails either extended up to the moving group in five-dimensional phase space, for example moving group HSC 759 near Melotte 111 and moving group HSC 2408 near Platais 10. After we analysed the overdensities and other observables mentioned above, these phase space interferences were removed. A deeper analysis with $N$-body simulations, spectroscopic and time-series data (for stellar gyro-chronology, as the proxy of the actual age) may reveal the kinematic and morphological connections between the OCs and associated moving groups better. These associations may be physical and may be examples of incomplete post-gas-expulsion coalescence of the clusters discussed by \citet{2024A&A...691A.293Zhou}.

\subsection{Luminosity function} 
\label{sec: LF}

\begin{figure}[h]
    
    \includegraphics[width=1\linewidth, height =1\linewidth]{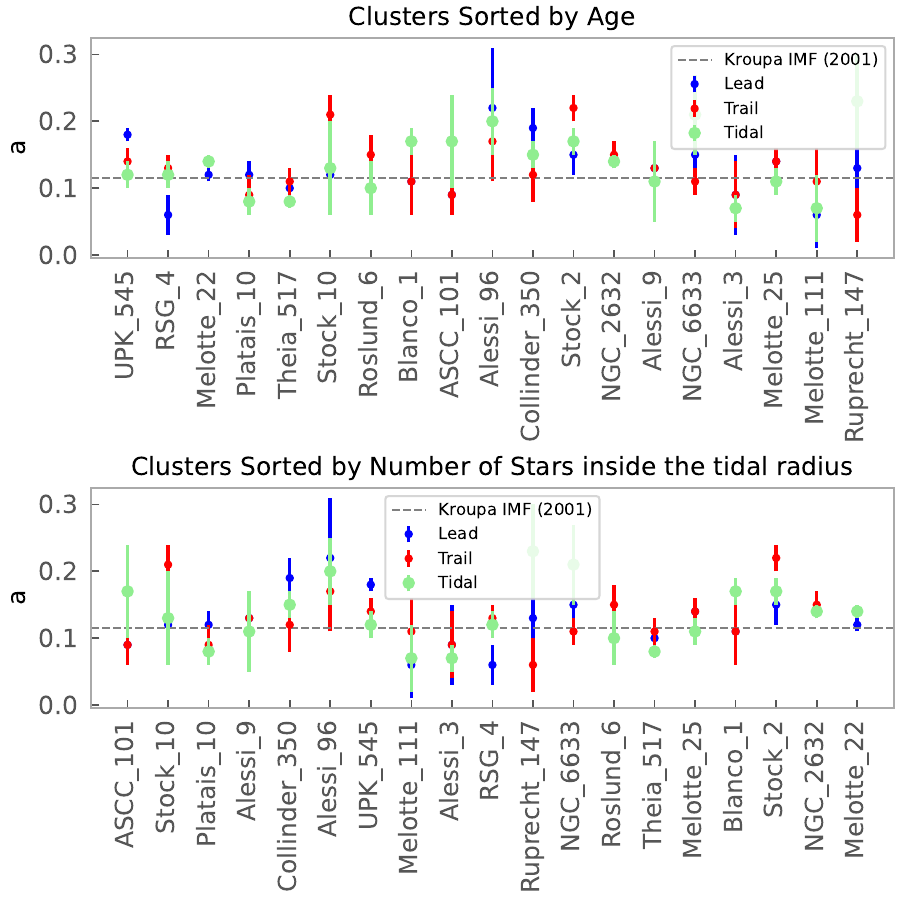}
    \caption{Luminosity function slopes for all the clusters in the region inside the tidal radius (green), the leading tail (blue), and the trailing tail (red). The grey line represents the nominal slope for the Gaia G band (0.115) for a cluster with a \citet{2001MNRAS.322..231KroupaIMF} initial mass function. }
    \label{fig: LF_slopes_img}
\end{figure}

The luminosity functions (LFs) for each cluster and its tidal tails were made using the corrected samples (see Section~\ref{sec: Corrected Samples}). They are shown in panel (h) of Figs.~\ref{fig:ASCC_101} to~\ref{fig:UPK_545}. For comparison purposes, we calculated the LFs of the region inside the tidal radius, the leading tail, and the trailing tail with the same $M_G$ bin size of 1 mag. 
The faintest one or two bins for all the clusters in all the three regions had fewer counts (likely due to observational incompleteness and stellar evaporation). They were therefore not included in the straight-line fit using the following relation:
\begin{equation}
    log_{10} N \propto a \times M_{G},
\end{equation}
where $N$ is the number of stars in each magnitude bin, and $a$ is the slope of the LF. Figure~\ref{fig: LF_slopes_img} shows that the slopes of the LFs in all the three regions are similar and are consistent with the observed value of $\approx$0.115 for a cluster with a canonical \citet{2001MNRAS.322..231KroupaIMF} stellar initial mass function.

\subsection{S-shaped tails that contradict known theory }
For 4 of the 19 OCs (Melotte 22, UPK 545, RSG 4 and Alessi 9) with clear evidence of tidal tails, we found S-shaped tidal tails, in which the leading tail was tilted towards the Galactic anti-centre. This disagrees with the theoretical models of tails, for example \citep{2020A&A...640A..84Dinnbier,2020MNRAS.497..536Wang,2021A&A...647A.137Jerabkova}, who predicted that the leading tail is tilted towards the Galactic centre. It also contradicts known theory. The comparisons of the basic parameters between the clusters and their respective inverted tails similar to the verification of an association of a moving group (see Section~\ref{sec: Moving Groups}) are strikingly consistent. In some cases, a nearby cluster may lie outside the selected phase space of the cluster. Some part of the extended tail can still be inside in the phase space, however. We confirmed for the inverted tails that they are not the tidal tails of such another nearby OC that is by chance aligned with the cluster and is incorrectly detected.  

We discuss a possible physical reasons of the formation of S-shaped tails. As tested to some extent and pointed out by \citet{2021A&A...647A.137Jerabkova}, the initial spin angular momentum of the cluster has a significant effect on the formation of tails and properties such as the mass and number of stars in both tails. A cluster that spins in the same direction as its orbital motion develops longer and more massive tails than a cluster without initial spin angular momentum, but the phase-space depopulation via the formation of tidal tails for the systems with non-zero initial spin is only poorly understood. A number of simulations of clusters that rotate with different initial spins in the direction of their orbital angular momentum and opposite to it need to be analysed to study this further. This might show the missing links between initially rotating clusters and the orientation of their tails.

Another reason for the inverted tails might be dynamical interactions of the cluster and tails with another nearby cluster or the Galactic disk. Several star clusters that form in the same molecular cloud may influence each other's tails. Stars that escape into the tails of a cluster may be subject to the gravitational potential of a neighbouring and comparatively massive cluster, which might be stronger than the Galactic tidal field. This would tilt the tails away from the Galactic centre. This possibility must be thoroughly tested using sets of simulations in which multiple neighbouring clusters from the same molecular cloud initially evolve together for a brief period of time at least.

\begin{comment}
    
\vik{Refactor the code and make a documented set of jupyter notebooks which can be used to do the cp/sccp. Also mention any manual steps taken as comments. I know the current version of the code is not publishable, but you need to start the process so that you can write publishable code in later projects.}
\end{comment}

\section{Conclusions} \label{sec:conclusions}

\begin{itemize}
    \item We analysed 21 nearby clusters older than 100 Myr with a minimum of 100 members using the CP (and the new SCCP) method to identify their tidal tails if they exist. We found tidal tails in 19 clusters and a stellar halo for one cluster. 
    \item The tails have a tip-to-tip span of 20-200 pc with 30-700 tail members. These statistics have different biases in the tracing of the tails due to some factors such as their different masses, distances, and orientation of the tails in $X-Y$ coordinates, however. We therefore suggest using the $M_{G}$ -corrected values for a comparison of the populations. 
    \item The analysis in the spherically rotated tail-aligned sky coordinate system ($\phi_1$, $\phi_2$) showed that the tail-aligned proper motions are mostly along the tail elongation and point towards the leading tail, consistent with expectations. The deviation seen in a few tails arises from a misalignment of the galactocentric velocities with the three-dimensional morphology of the tails.
    \item The RVs of the tails are consistent with a mean radial velocity of the OC, with the trailing tail RVs being slightly higher than the leading tail.
    \item The LFs of the tails in the vicinity of the clusters are similar to those of the cluster. The slope of the LFs in the region inside the tidal radius, the leading tail, and the trailing tail are mostly consistent, but the uncertainties are larger. 
    \item The tidal shapes are expected to be \reflectbox{S}-shaped, with the tilt of the leading tails being towards the Galactic centre, while the trailing tails tilt away from the Galactic centre. However, 4 out of 19 clusters show S-shaped tails. 
    The CMD (proxy for the stellar age), energy, and angular momentum are all consistent with a co-eval and co-moving group of stars.
    Interactions with a clumpy Galactic potential, an initial spin of the clusters, or a preferred detection of co-moving stars from a common molecular cloud might be the cause of this. 
    The further analysis of these theory-defying tails using computer simulations of clusters and their tidal tails is required for a better understanding.
\end{itemize}

In summary, while it was thought that the tidal tails of the open stars clusters ought to be dynamically and kinematically fairly simple structures (\citealt{2020A&A...640A..84Dinnbier,2021A&A...647A.137Jerabkova,2010MNRAS.401..105K,2009MNRAS.392..969Just}), the actual real tidal tails appear to show features that are not understood at present and are certainly unexpected \citep[e.g.][]{2022MNRAS.517.3613Kroupa,2024ApJ...970...94Kroupa}.

\section*{Data availability}
The catalogue of cluster and tidal tail members is available in electronic form at the CDS via anonymous ftp to \href{https://cdsarc.u-strasbg.fr}{cdsarc.u-strasbg.fr (130.79.128.5)} or via \href{http://cdsweb.u-strasbg.fr/cgi-bin/qcat?J/A+A/}{http://cdsweb.u-strasbg.fr/cgi-bin/qcat?J/A+A/}.

The plots from our analysis for all the clusters in Table~\ref{Table: Tidal tails} is available on Zenodo \href{https://doi.org/10.5281/zenodo.14651597}{https://doi.org/10.5281/zenodo.14651597}.

\begin{acknowledgements}
    % We thank the anonymous referee for constructive comments. 
    Authors of this paper acknowledge anonymous referee for their comments which improved the quality of this work significantly.VJ thanks the Alexander von Humboldt Foundation for their support.
    PK acknowledges support through the DAAD Eastern European Bonn-Prague Exchange Programme. 
    We thank S. Bhattacharya (IUCAA, India) for sharing their catalogue.
    We thank K. Malhan (MPA, Germany) for help with the orbital calculations.
    This work has made use of data from the European Space Agency (ESA) mission {\it Gaia} (\url{https://www.cosmos.esa.int/gaia}), processed by the {\it Gaia} Data Processing and Analysis Consortium (DPAC, \url{https://www.cosmos.esa.int/web/gaia/dpac/consortium}). Funding for the DPAC has been provided by national institutions, in particular the institutions participating in the {\it Gaia} Multilateral Agreement.
    The following software were used in this work: Astropy \citep{astropy:2022}; matplotlib; Gala \citep{2017JOSS....2..388Price}; Galpy\citep{Bovy2015ApJS..216...29B}; \citep{2007CSE.....9...90Hunter}; SciPy \citep{2020SciPy-NMeth}; NumPy \citep{2020Natur.585..357Harris}; TOPCAT \citep{2005ASPC..347...29TOPCAT}.
\end{acknowledgements}
\bibliographystyle{aa} % style aa.bst
\bibliography{references}

\begin{appendix}

\section{Supplementary query and tables}\label{Appendix:Appendix_B}

\subsection{ADQL query} \label{sec:adql}

ADQL query for selecting the Gaia DR3 data

\noindent\rule[0.5ex]{\linewidth}{0.5pt}
{\footnotesize {\fontfamily{pcr}\selectfont
\noindent 
\textcolor{magenta}{select} TOP 10000000
\\dr3.source\_id, dr3.ra, dr3.dec, dr3.parallax, dr3.pmra, dr3.pmdec,  dr3.bp\_rp, dr3.phot\_g\_mean\_mag, \\ 1000 * dr3.parallax\_error / \textcolor{magenta}{power}(dr3.parallax, 2) AS Dist\_error
\\\textcolor{magenta}{from} gaiadr3.gaia\_source AS dr3 \\
\textcolor{magenta}{where}
    parallax > 2 \\
    \textcolor{magenta}{and} pmra \textcolor{magenta}{IS NOT NULL} \\
    \textcolor{magenta}{and} pmdec \textcolor{magenta}{IS NOT NULL}  \\
    \textcolor{magenta}{and} parallax \textcolor{magenta}{IS NOT NULL}  \\
    \textcolor{magenta}{and} 1000 * dr3.parallax\_error / 
        \textcolor{magenta}{power} (dr3.parallax, 2) < 10 \\
    \textcolor{magenta}{and} (astrometric\_excess\_noise < 1 
        OR astrometric\_excess\_noise\_sig < 2)\\
    \textcolor{magenta}{and} ruwe < 1.4\\
    \textcolor{magenta}{and} phot\_g\_mean\_mag \textcolor{magenta}{IS NOT NULL} \\
    \textcolor{magenta}{and} phot\_bp\_mean\_mag \textcolor{magenta}{IS NOT NULL} \\
    \textcolor{magenta}{and} phot\_rp\_mean\_mag \textcolor{magenta}{IS NOT NULL} \\

\noindent\rule[0.5ex]{\linewidth}{0.5pt}
}}

\subsection{Tables}

\begin{table}[h]
\label{Table: phi_org_pole}
\caption{Pole and origin in equatorial coordinates used for every cluster for the transformations from ($\alpha$, $\delta$) to ($\phi_1$,$\phi_2$) and ($\mu_{\alpha} cos \delta $, $\mu_{\delta}$) to ($\mu_{\phi_{1}} cos \phi_2 $, $\mu_{\phi_{2}}$) (see section \ref{sec: phi1-phi2}).} 
\centering
\begin{tabular}{c c c c c}

\hline
Cluster  &$\alpha_{\rm{pole}}$   & $\delta_{\rm{pole}}$ & $\alpha_{\rm{org}}$ & $\delta_{\rm{pole}}$ \\

         & (\rm{deg}) & (\rm{deg}) & (\rm{deg}) & (\rm{deg}) \\
\hline
Alessi\_3            & 120           & 43            & 108.993         & -46.233         \\
Alessi\_9            & 212           & 28            & 265.951         & -46.954         \\
Alessi\_96           & 193           & 31            & 75.501          & 37.506          \\
ASCC\_101            & 170           & 33            & 288.354         & 36.310          \\
Blanco\_1            & 57            & 44            & 53.057          & -29.883         \\
Collinder\_350       & 153           & 86            & 267.038         & 1.537           \\
Melotte\_111         & 221           & -60           & 186.247         & 25.953          \\
Melotte\_22          & 162           & 31            & 56.626          & 24.095          \\
Melotte\_25          & 176           & 43            & 67.469          & 16.920          \\
NGC\_2632            & 39            & 4             & 129.953         & 19.589          \\
NGC\_6633            & 181           & 41            & 276.890         & 6.635           \\
Platais\_10          & 11            & -30           & 205.692         & -59.348         \\
RSG\_4               & 343           & -21           & 288.020         & 56.871          \\
Roslund\_6           & 13            & -26           & 307.285         & 39.889          \\
Ruprecht\_147        & 203           & 13            & 289.053         & -16.465         \\
Stock\_10            & 16            & -25           & 84.801          & 37.754          \\
Stock\_2             & 189           & 28            & 33.834          & 59.531          \\
Theia\_517           & 188           & 32            & 322.768         & 48.293          \\
UPK\_545             & 25            & -7            & 130.795         & -58.687         \\
\hline
\end{tabular}

\end{table}

\newgeometry{margin= 0.5 in}
\begin{landscape}
\begin{table}
\centering
\caption{Open clusters selected in this work }
\label{Table: Clusters_basics_params} 
\begin{tabular}{c c c c c c c c c c c}  
\hline\hline           
Name & Other names & $\alpha_c $ & $\delta_c$ & $\mu_{\alpha} \cos{\delta}$ & $\mu_{\delta}$& $\omega$&  $\alpha_{cp} $   & $\delta_{cp} $ & log\textsubscript{10}$(\frac{age}{yr})$ & $r_t$  \\ 
     &              & $(\rm{deg})$ & $(\rm{deg})$ & $(\rm{mas \ yr^-1})$ & $(\rm{mas \ yr^-1})$ & $(\rm{mas})$ &
$(\rm{deg})$ & $(\rm{deg})$ & &  (pc) \\   
\hline     
    ASCC 101 & Alessi J1913.6+3619, & 288.345   & 36.360  & 0.967     & 1.281     & 2.51 & 280.4 $\pm$ 2.0 & 27.0 $\pm$ 2.7  & 8.3 & 5.69$^a$ \\
            & LeDrew 6,MWSC 3070,Theia 445 & \\
\hline            
    Alessi 3 & MWSC 1157,Theia 1011 & 109.187   & -46.680 & -9.793    & 11.916    & 3.58 & 45.4 $\pm$ 3.0 & 48.1 $\pm$ 2.3 &  8.8 & 7 $^b$\\
\hline
    Alessi 9 & MWSC 2670 & 265.642   & -46.744 & 9.831    & -9.151   &  4.78 & 228.4 $\pm$ 1.0 & -1.5 $\pm$ 1.8  & 8.6 & 5.5$^b$\\
\hline
    Alessi 96 & RSG 1,TRSG 1,Theia 380  & 75.485    & 37.487  & 0.971   & -12.561    & 3.05 & 77.7 $\pm$ 0.2   &  6.7 $\pm$ 4.0  &  8.3 & 10$^c$\\
\hline
    Blanco 1 & ESO 409-9,MWSC 7,OCL 43,zeta Scl  & 0.914     & -30.011 & 18.725     & 2.593    & 4.23 & 78.8 $\pm$ 4.5 & 1.8 $\pm$ 2.6 & 8.2 & 8.5$^d$\\
\hline
    CWNU 1084 & & 302.547   & 33.490  & 5.048    & -0.388   &  3.10 & 273.0 $\pm$ 1.9  & 31.5 $\pm$ 0.4 & 8.0& 8.1 $^b$ \\
\hline
    Collinder 350 &MWSC 2700,OCL 75,Theia 919 & 267.065   & 1.418   & -4.869    & -0.066    & 2.69 & 303.9 $\pm$ 7.5 & 1.9 $\pm$ 0.1  & 8.4 &  7.4 $^b$\\
\hline
    Melotte 111 & Collinder 256,Coma Berenices,MWSC 2020,OCL 558 & 186.017   & 26.423  & -12.155    & -8.855   & 11.71 & 276.4 $\pm$ 5.7 & 39.1 $\pm$ 1.5 & 8.8 & 6.9$^e$\\ 
\hline
    Melotte 22 & Collinder 42,Escorial 1,M45,MWSC 305, & 56.680   & 24.108  & 19.955 & -45.457    & 7.38 & 89.8 $\pm$ 0.6 & -44.7 $\pm$ 0.7 & 8.1 & 11.8$^d$\\
               & OCL 421,Pleiades,Theia 369 & \\
\hline
    Melotte 25 & Collinder 50,Hyades,MWSC 379 & 66.701  & 16.081  & 104.136 & -28.732   & 21.2 & 97.3 $\pm$ 0.2 & 6.9 $\pm$ 0.1 & 8.8 & 9.0$^k$\\
               & OCL 456,Taurus Moving Cluster & \\
\hline
    NGC 2632 & Collinder 189,MWSC 1527,M 44,  & 130.088   & 19.666  & -35.943   & -12.903    & 5.40 & 89.8 $\pm$ 0.5 & 1.6 $\pm$ 0.3  & 8.5 & 10.77$^f$\\               &Melotte 88,OCL 507,Praesepe  & \\
\hline    
    NGC 6475 & BH 254,Collinder 354,ESO 394-9,MWSC 2739,  & 268.476   & -34.828 & 3.080    & -5.366    & 3.58 & 254.1 $\pm$ 0.7 & -8.2 $\pm$ 1.6 & 8.2 & 14.1$^b$\\
               & Melotte 183,OCL 1028,Theia 692 & \\
\hline
    NGC 6633 & Collinder 380,MWSC 2917, & 276.805   & 6.559   & 1.204    & -1.804     & 2.54 & 265.6 $\pm$ 3.5 & 21.6 $\pm$ 3.4  & 8.7 & 10.2$^b$\\
             & Melotte 201,OCL 90,Theia 924   & \\ 
\hline               
    Platais 10 &HIP 67014 Group,MWSC 2150 & 205.279   & -59.156 & -29.312   & -10.788    & 4.13 & 271.7 $\pm$ 0.1   & -3.4 $\pm$ 0.1 &  8.1 & 10 $^c$\\    
\hline
    RSG 4 & TRSG 4,Theia 520,UBC 1  & 287.696   & 56.523  & -2.489     & 3.734    & 3.09 & 298.4 $\pm$ 0.8 & 45.0 $\pm$ 1.2 &   8.0& 17.5$^g$\\    
\hline    
    Roslund 6 & RSG 6,TRSG 6 & 307.593   & 40.356  & 5.950     & 2.130    & 2.83 & 250.3 $\pm$ 6.6   & 3.5 $\pm$ 6.3 & 8.2 & 7.5 $^a$\\

\hline    
    Ruprecht 147& MWSC 3078,NGC 6774,OCL 65,Theia 1531  & 289.141   & -16.269 & -0.949   & -26.672    & 3.28 & 286.3 $\pm$ 0.3 & -59.0 $\pm$ 2.6 &  8.9 &  12.11$^h$\\
\hline    
    Stock 10 &Theia 609 & 84.742    & 37.808  & -3.150    & -0.428    & 2.79 & 117.2 $\pm$ 3.2 & 36.8 $\pm$ 0.5  & 8.1 & 5.04$^j$\\
\hline    
    Stock 2   & Theia 524  & 33.850   & 59.577  & 15.829 & -13.713  & 2.68 & 80.2 $\pm$ 0.7 & 3.8 $\pm$ 1.6 &  8.5 & 22.65$^i$\\
\hline
    Theia 517 & Collinder 438,M 39, MWSC 3521,Melotte 236,& 322.831   & 48.349  & -7.400   & -19.704    & 3.37 & 289.8 $\pm$ 2.5  & -50.2 $\pm$ 3.9   & 8.1 & 6.7$^d$\\
             & NGC 7092,OCL 211,Theia 822  & \\
\hline    
    UPK 545 &  & 130.975   & -58.784 & -8.417     & 2.531    & 2.95 & 100.9 $\pm$ 3.1 & -48.0 $\pm$ 2.0 & 8.0 & 10$^c$\\

\hline                                          
\end{tabular}
\vspace{0.5 cm}
  \par \textit{} Columns 1 \& 2 : Name(s) of the OC from \citetalias{2024arXiv240305143H}; Columns 3 \& 4 : Sky-coordinates from \citetalias{2024arXiv240305143H}; Columns 5 \& 6: Proper motions from \citetalias{2024arXiv240305143H} ; Column 7: Parallax from \citetalias{2024arXiv240305143H}; Columns 8 \& 9: Convergent-point Coordinates found in this work (see the Section~\ref{sec: CP}); Column 10: Logarithm the of age of the OC from \citetalias{2024arXiv240305143H}; Column 11: Tidal radius in parsec from various sources,
  $^a$\citet{10.1093/mnras/stac2906},
  $^b$\citet{2022ApJ...931..156Pang},
  $^c$ Canonical Assumption,
  $^d$\citet{2019A&A...621L...3Meingast}
  $^e$\citet{2019ApJ...877...12Tang}, 
  $^f$\citetalias{2019A&A...627A...4Roser_Praesepe}, 
  $^g$\citetalias{2024arXiv240305143H},
  $^h$\citet{2019AJ....157..115Yeh}
  $^i$\citet{2021AJ....162..171Ye}, 
  $^j$\citet{2023yCat..36720187Just}
  $^k$\citet{2011A&A...531A..92RoserHyades_Not_tails}.
\end{table}

\begin{table}
\centering
\caption{Statistics of the Tidal tails we found in this work}. 
\label{Table: Tidal tails}    
\begin{tabular}{c c c c c c c c c c c c }      
\hline\hline                       
Name & N\textsubscript{Tidal} & N\textsubscript{Leading} & N\textsubscript{Trailing} & N\textsubscript{Tidal,Corr}& N\textsubscript{Leading,Corr}&  N\textsubscript{Trailing,Corr}& 
$M_{G,Lim}$ & $p$\textsubscript{cont,max} & $a$  & $b$ & $r\textsubscript{lim}$ \\ 

 &  &  &   & &  &  & (mag) &  & (km \ s$^{-1}$)  & (km \ s$^{-1}$) & (pc) \\ 

\hline     
ASCC 101 & 30 & 33 & 81 & 26 & 26 & 25 & 8.173   & 0.097 & 2.4 & 1.8 & 17\\

Alessi 3$^*$ & 95 & 91 & 47 & 75 & 36 & 32 & 10.114 & 0.090 & 1.6 & 1.0 & 20\\

Alessi 9 & 52 & 148 & 185 & 43 & 82 & 115 & 10.464 &  0.070 & 2.0 & 1.0 & 17 \\    

Alessi 96 & 65 & 110 & 60 & 50 & 66 & 42 & 8.724  &0.102 & 3.0 & 1.5 & 20\\

Blanco 1 & 404 & 68 & 82 & 270 & 42 & 41 & 10.612 & 0.089 & 4.0 & 2.0 & 15\\

Collinder 350$^*$ & 53 & 52 & 130 & 41 & 42 & 61 & 7.875 &  0.090  & 2.5 & 1.5 & 23 \\

Melotte 111$^*$ & 85 & 72 & 55 & 79 & 58 & 47 & 12.935   & 0.093 & 1.2 & 1.2 & 20  \\

Melotte 22 & 1176 & 565 & 288 & 1144 & 411 & 272 & 13.17 & 0.098 & 1.5 & 0.9 & 20 \\

Melotte 25$^*$ & 312 & 423 & 334 & 305 & 349 & 329 & 14.759 &0.080 & 2.5 & 0.8 & 15 \\

NGC 2632$^*$ & 973 & 308 & 147 & 856 & 168 & 132 & 11.928 & 0.078 & 2.0 & 1.0 &17\\

NGC 6633 & 166 & 87 & 105 & 147 & 71 & 76 & 7.901 & 0.090 & 2.5 & 1.8 & 18   \\

Platais 10$^*$ & 48 & 89 & 67 & 38 & 64 & 37 & 10.669  & 0.088 & 3.5 & 0.5 & 17\\

RSG 4 & 110 & 64 & 247 & 77 & 49 & 73 & 9.117  & 0.106 & 2.5 & 1.8 & 20\\

Roslund 6 & 191 & 43 & 123 & 147 & 34 & 62 & 8.629 & 0.048 & 1.5 & 1.1 & 17\\

Ruprecht 147$^*$ & 127 & 69 & 139 & 113 & 58 & 49 & 9.055  & 0.100 & 3.8 & 1.2 & 20\\

Stock 10$^*$ & 41 & 81 & 73 & 34 & 53 & 56 & 8.089 & 0.091 & 2.1 & 1.7 & 25\\

Stock 2 & 883 & 118 & 277 & 771 & 95 & 219 & 8.609 & 0.092 & 7.0 & 2.1 & 14\\

Theia 517$^*$ &  211 & 83 & 195 & 145 & 71 & 55 & 9.707 & 0.206 & 3.0 & 1.5 & 18\\

UPK 545 & 83 & 382 & 134 & 72 & 118 & 102 & 9.207 & 0.085 &2.0 & 1.5 & 17\\

\hline                                          
\end{tabular}
\vspace{0.5 cm}
  \par \textit{} Column 1 : Name of the OC from \citetalias{2024arXiv240305143H}, $^*$ signifies the statistics for CP + SCCP joint sample (see Section~\ref{sec: SCCP}); Columns 2 - 4 : Number of stars inside the tidal radius, the leading tail and the trailing tail ; Columns 5 - 7: Number of stars inside the tidal radius, the leading tail and the trailing tail for the corrected sample (see the Section~\ref{sec:results})  ; Column 8: Maximum absolute G magnitude for the correction (see Section~\ref{sec:results}); Column 9: Maximum contamination probability for k = 6 (see Section~\ref{sec: CP}); Columns 10 - 12: Optimum neighbour finding parameters (see Section~\ref{sec:finding_overdensity}).

\end{table}

\begin{table}[h!]
  
  \caption{Moving Groups (MvGr) associated with OCs and their tails found in this work.} 
  \label{Table: Moving_Groups} 
  \begin{tabular}{c c c c c c c c c c}      
  
  \hline\hline                       
   Name\_OC & Name\_{MvGr} & log\textsubscript{10}$(\frac{age}{yr})$\_OC & log\textsubscript{10}$(\frac{age}{yr})$\_{MvGr} & $V_X$\_OC (km s$^{-1})$& $V_Y$\_OC (km s$^{-1})$& $V_Z$\_OC (km s$^{-1})$& $V_X$\_{MvGr} (km s$^{-1})$ & $V_Y$\_{MvGr} (km s$^{-1})$ & $V_Z$\_{MvGr} (km s$^{-1})$\\ 
   \hline     
   Melotte 22 & HSC 1580 &  8.1 & 8.1 &  6.406 & 217.2 & -6.3 & 4.2 & 217.9 & -7.1  \\ 
   Melotte 22 & UPK 303 &  8.1 & 8.0 &  6.406 & 217.2 & -6.3 & 6.5 & 217.1 & -3.7  \\
   NGC 2632 & HSC 1244\textsuperscript{1} &  8.5 & 8.1  & -29.2 & 225.5 & -2.3 & -25.0 & 228.1 & -5.1  \\
       
   \hline
  \end{tabular}
  \vspace{0.5 cm}
  \par \textit{} First two columns are name of the OC and the moving group from \citetalias{2024arXiv240305143H}. Columns 3 and 4 are ages respectively. Columns 5 to 7 and 8 to 10 are Galactic velocities in $X$, $Y$ and $Z$ directions respectively. 
  1. A positional cross-match performed between tidal tails of Praesepe from \citetalias{2019A&A...627A...4Roser_Praesepe} and \citetalias{2024arXiv240305143H} shows that moving group HSC 1244 was also detected within the tidal tails of Praesepe by \citetalias{2019A&A...627A...4Roser_Praesepe}.

\end{table}

\end{landscape}
\restoregeometry

\newpage

\subsection{Appendix Figures of individual clusters and their description}\label{sec:Fig_Desc}

\begin{figure*}[h]
    
    \includegraphics[width=0.95\textwidth]{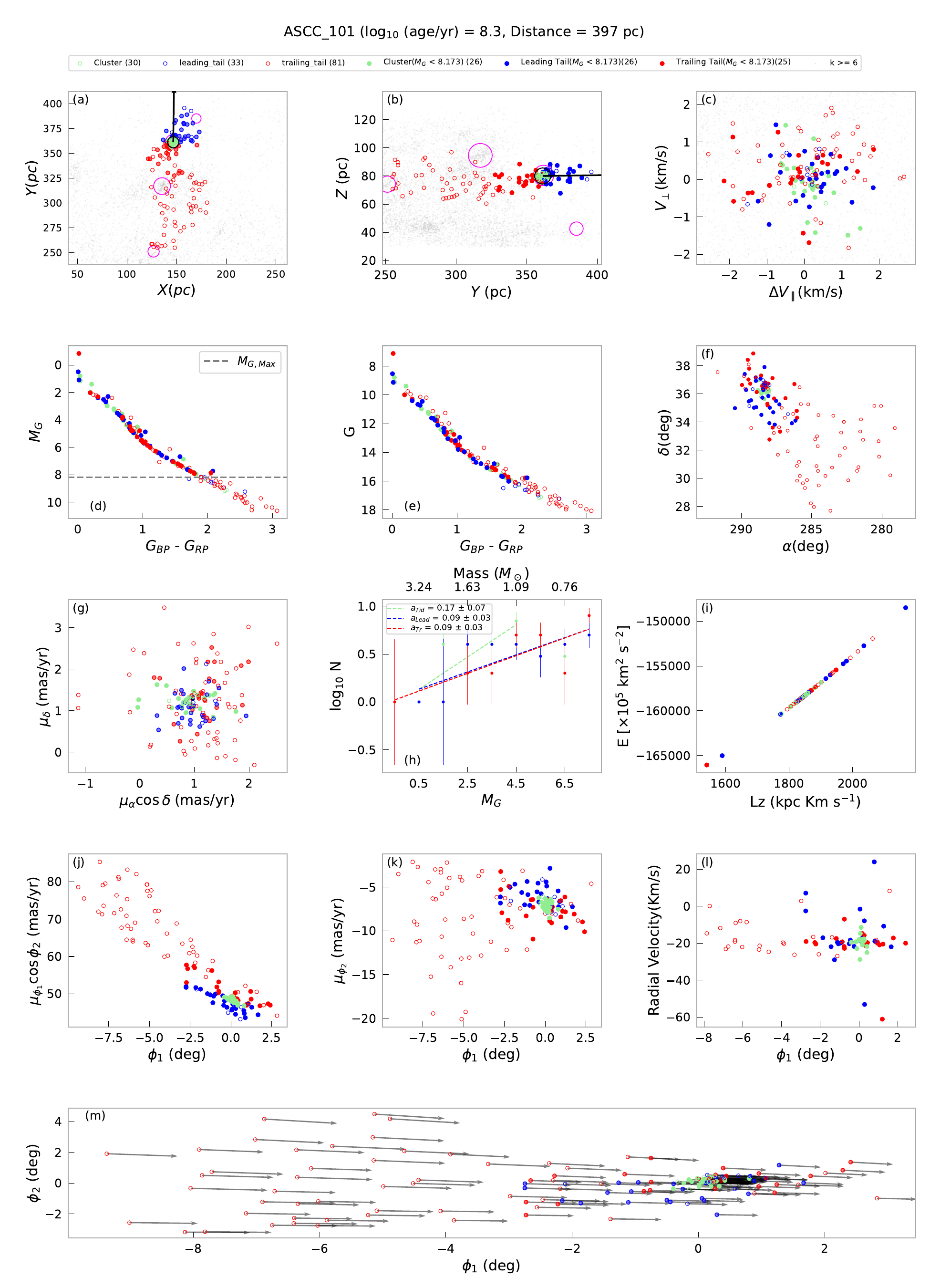}
    
    \caption{Results from our analysis for ASCC 101:
    The image description can be found in the Appendix~\ref{sec:Fig_Desc}}
    \label{fig:ASCC_101}
\end{figure*}

\begin{figure*}[h]
    \includegraphics[width=0.95\textwidth]{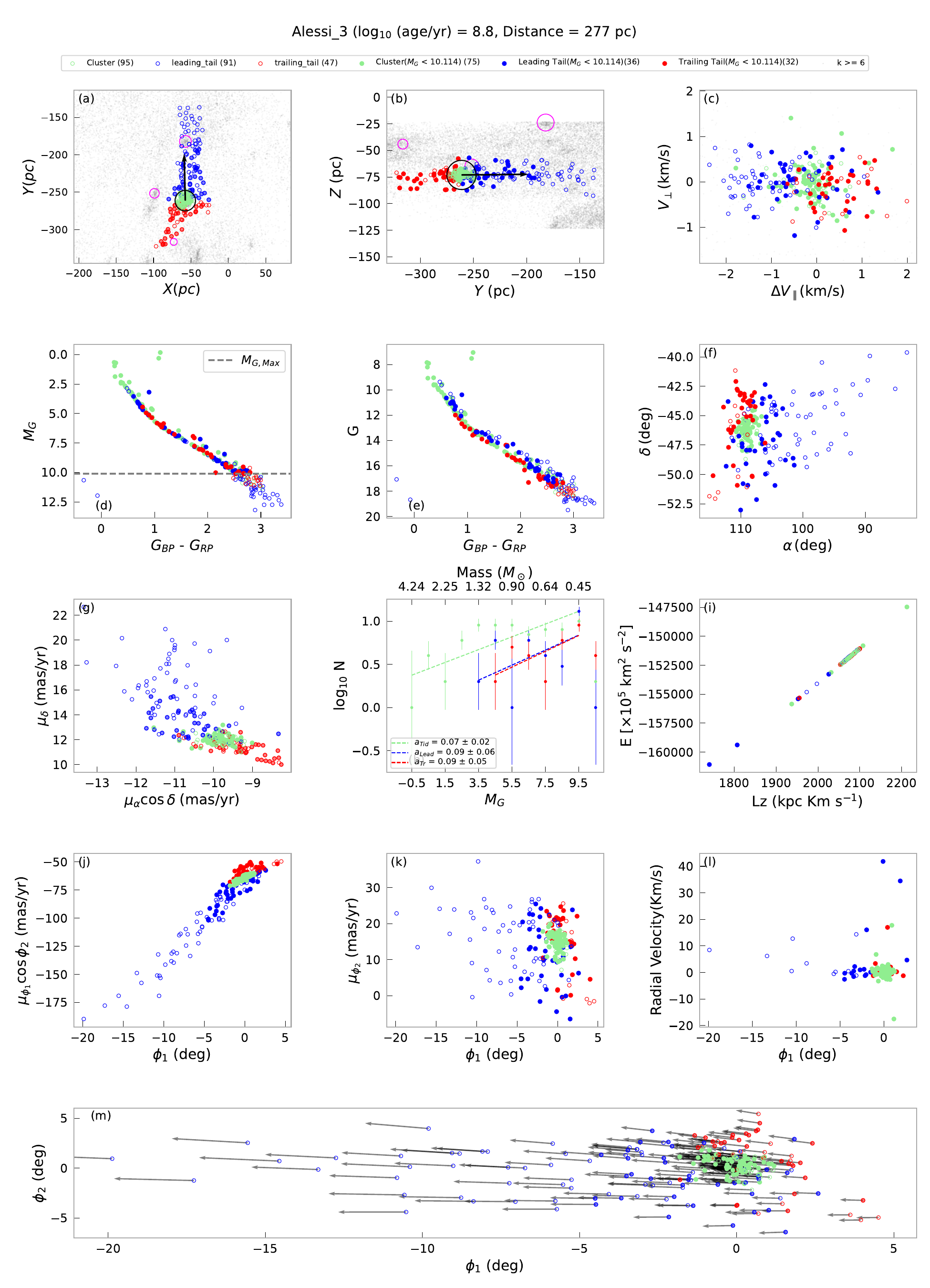}
    \caption{Same as the Figure~\ref{fig:ASCC_101} for Alessi 3}
    \label{fig:Alessi_3}
\end{figure*}

\begin{figure*}[h]
    \includegraphics[width=0.95\textwidth]{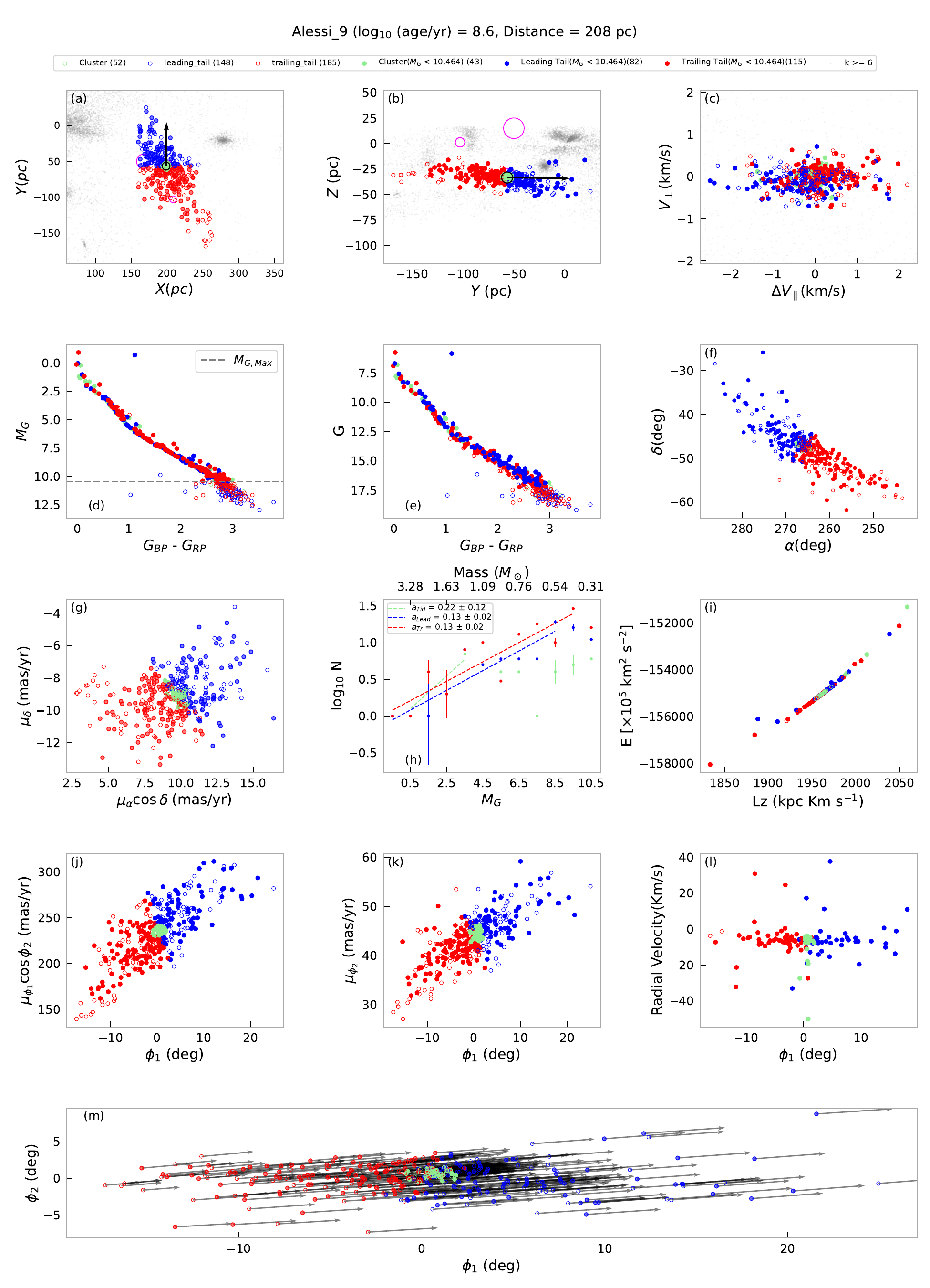}
    \caption{Same as the Figure ~\ref{fig:ASCC_101} for Alessi 9}
    \label{fig:Alessi_9}
\end{figure*}

\begin{figure*}[h]
    \includegraphics[width=0.95\textwidth]{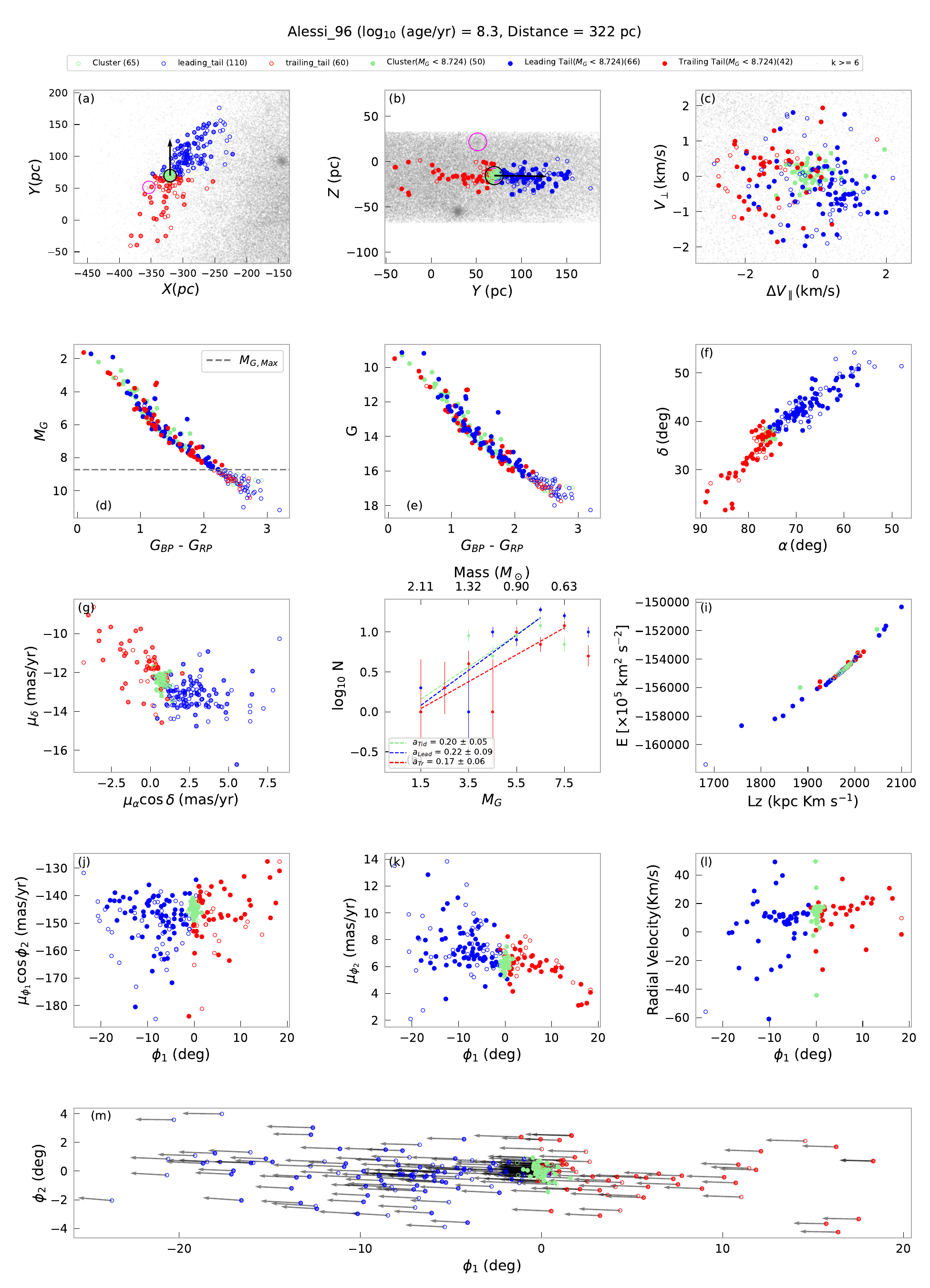}
    \caption{Same as the Figure~\ref{fig:ASCC_101} for Alessi 96}
    \label{fig:Alessi_96}
\end{figure*}

\begin{figure*}[h]
    \includegraphics[width=0.95\textwidth]{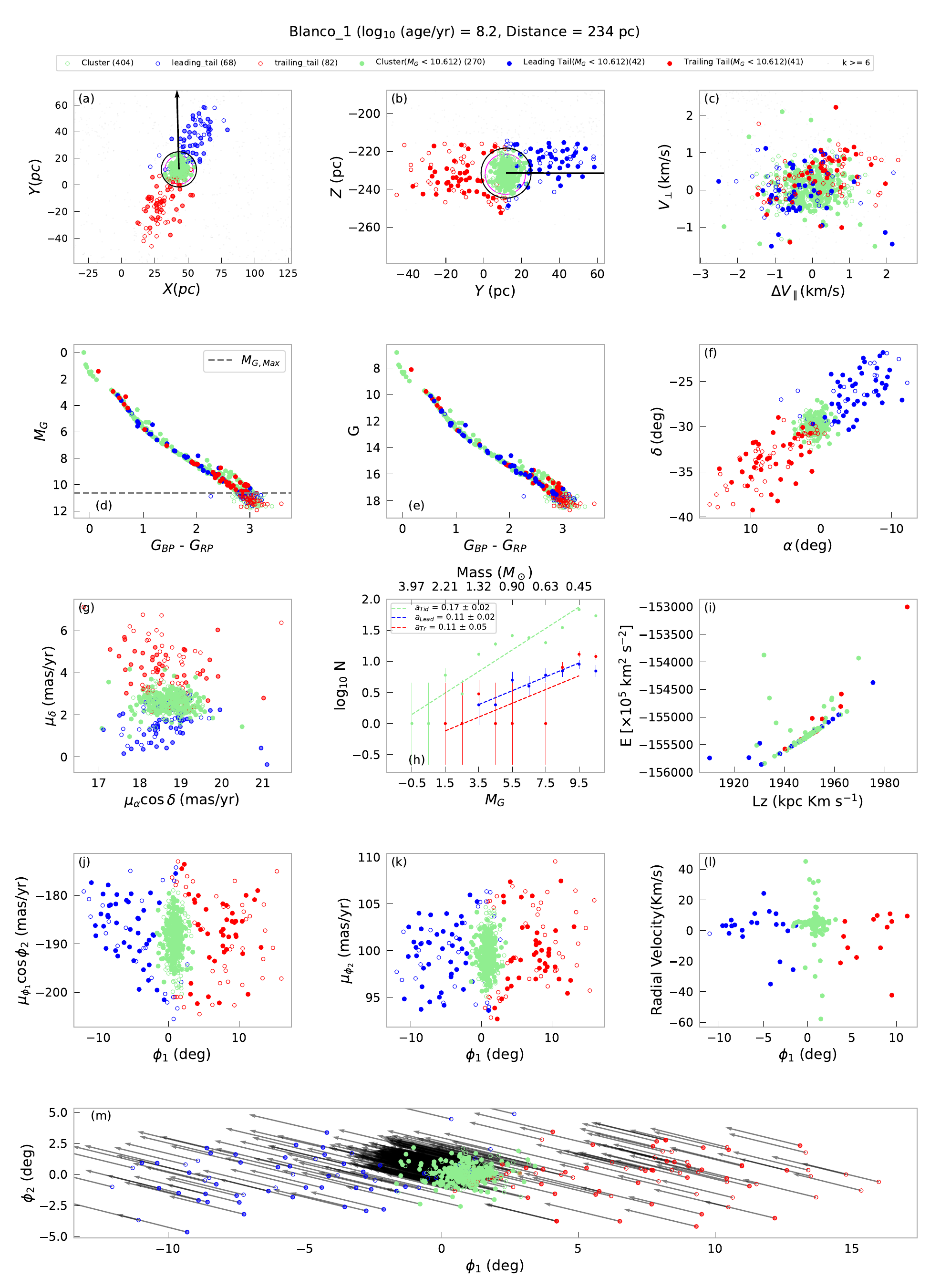}
    \caption{Same as the Figure~\ref{fig:ASCC_101} for Blanco 1}
    \label{fig:Blanco_1}
\end{figure*}

\begin{figure*}[h]
    \includegraphics[width=0.95\textwidth]{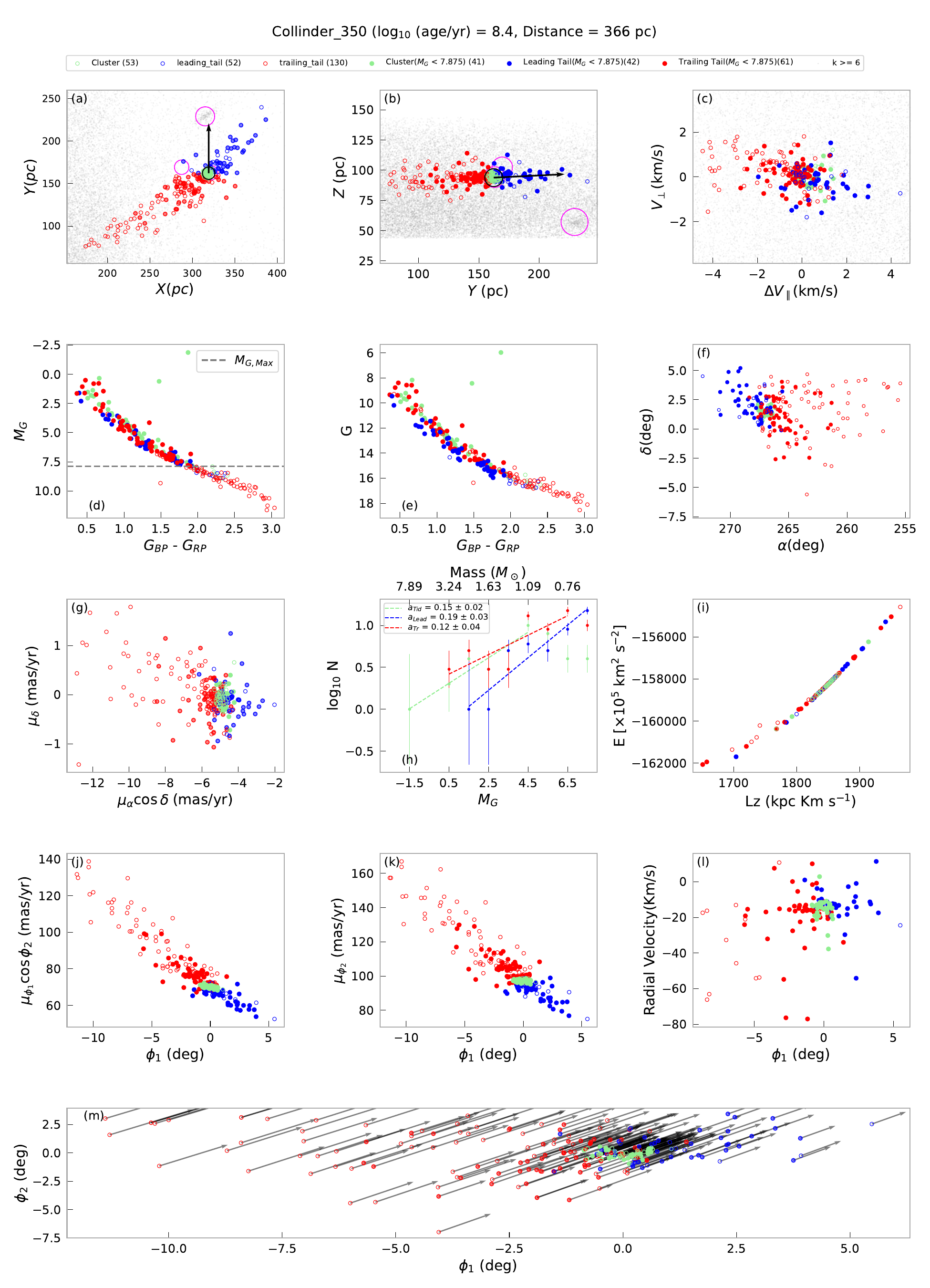}
    \caption{Same as the Figure~\ref{fig:ASCC_101} for Collinder 350}
    \label{fig:Collinder_350}
\end{figure*}

\begin{figure*}[h]
    \includegraphics[width=0.95\textwidth]{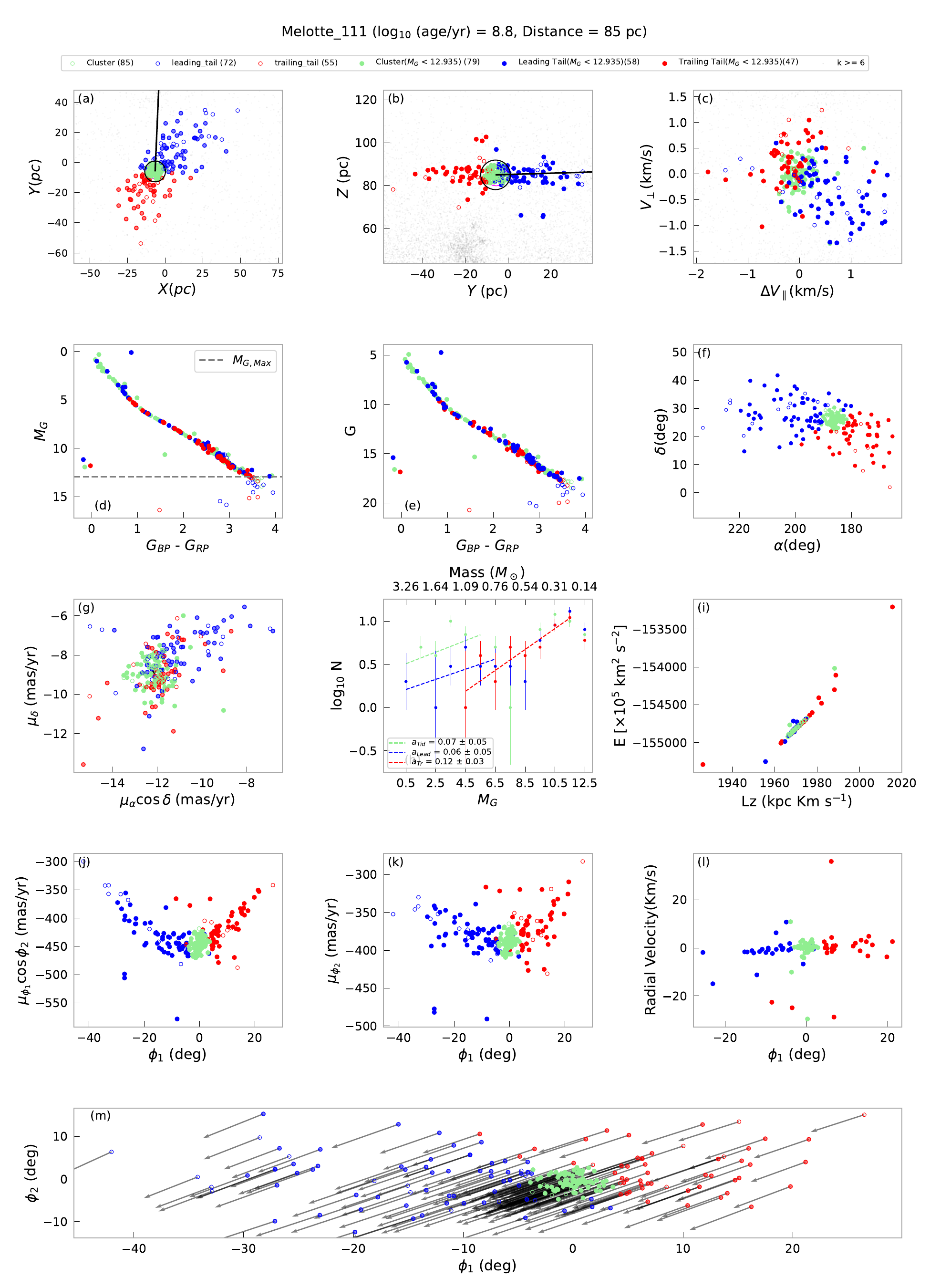}
    \caption{Same as the Figure~\ref{fig:ASCC_101} for Melotte 111}
    \label{fig:Melotte_111}
\end{figure*}

\begin{figure*}[h]
    \includegraphics[width=0.95\textwidth]{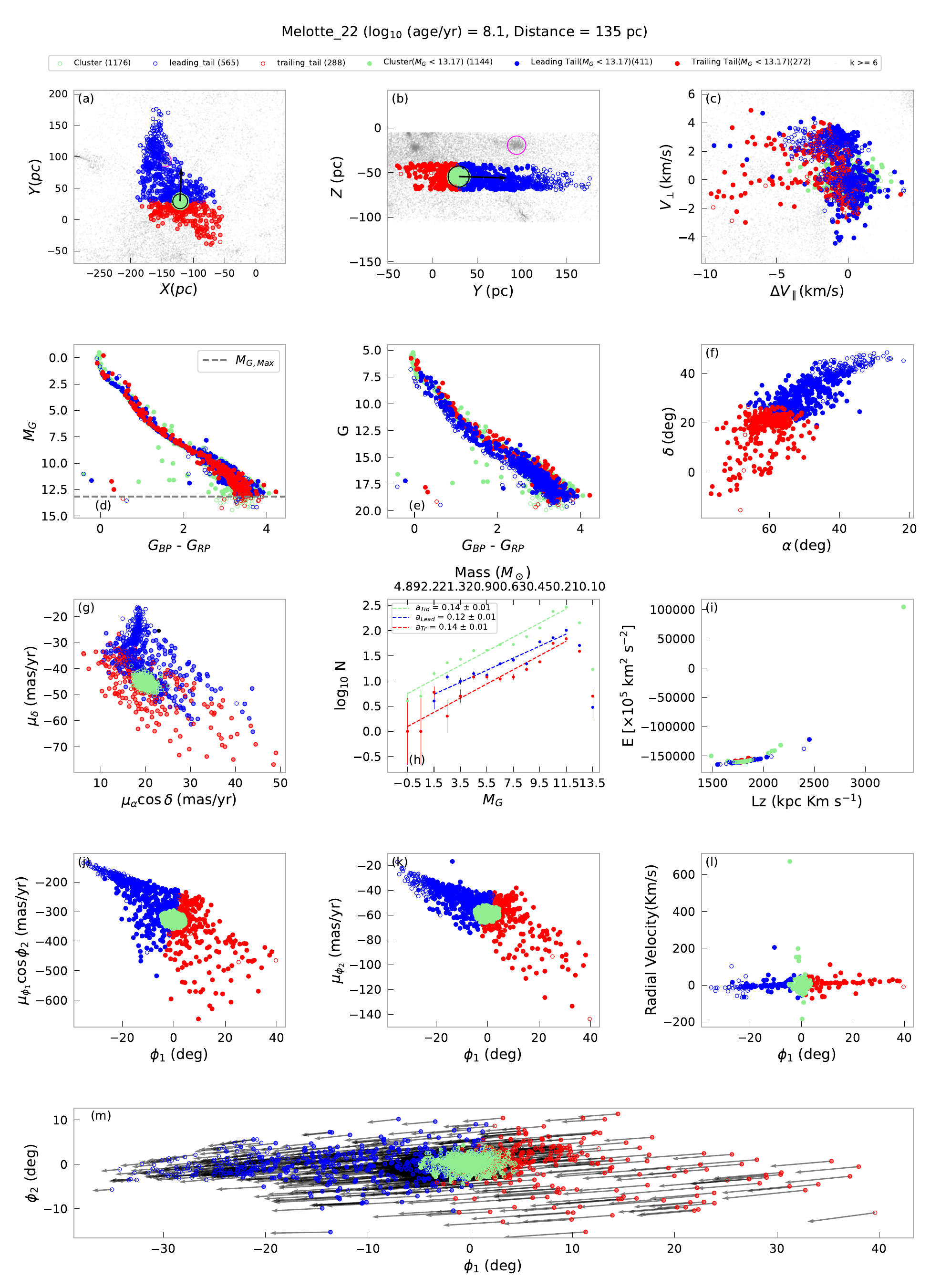}
    \caption{Same as the Figure~\ref{fig:ASCC_101} for Melotte 22}
    \label{fig:Melotte_22}
\end{figure*}

\begin{figure*}[h]
    \includegraphics[width=0.95\textwidth]{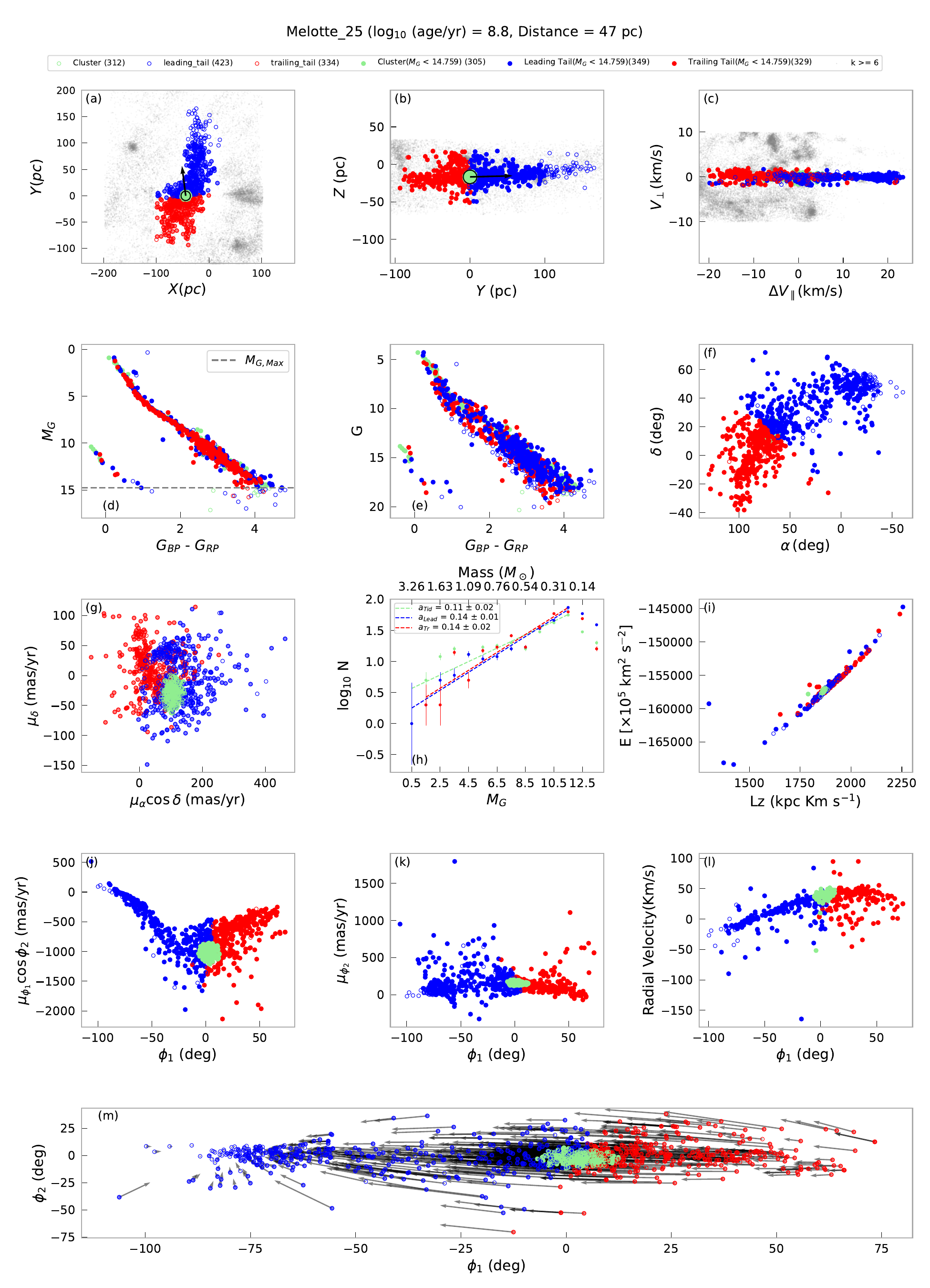}
    \caption{Same as the Figure~\ref{fig:ASCC_101} for Melotte 25}
    \label{fig:Melotte_25}
\end{figure*}

\begin{figure*}[h]
    \includegraphics[width=0.95\textwidth]{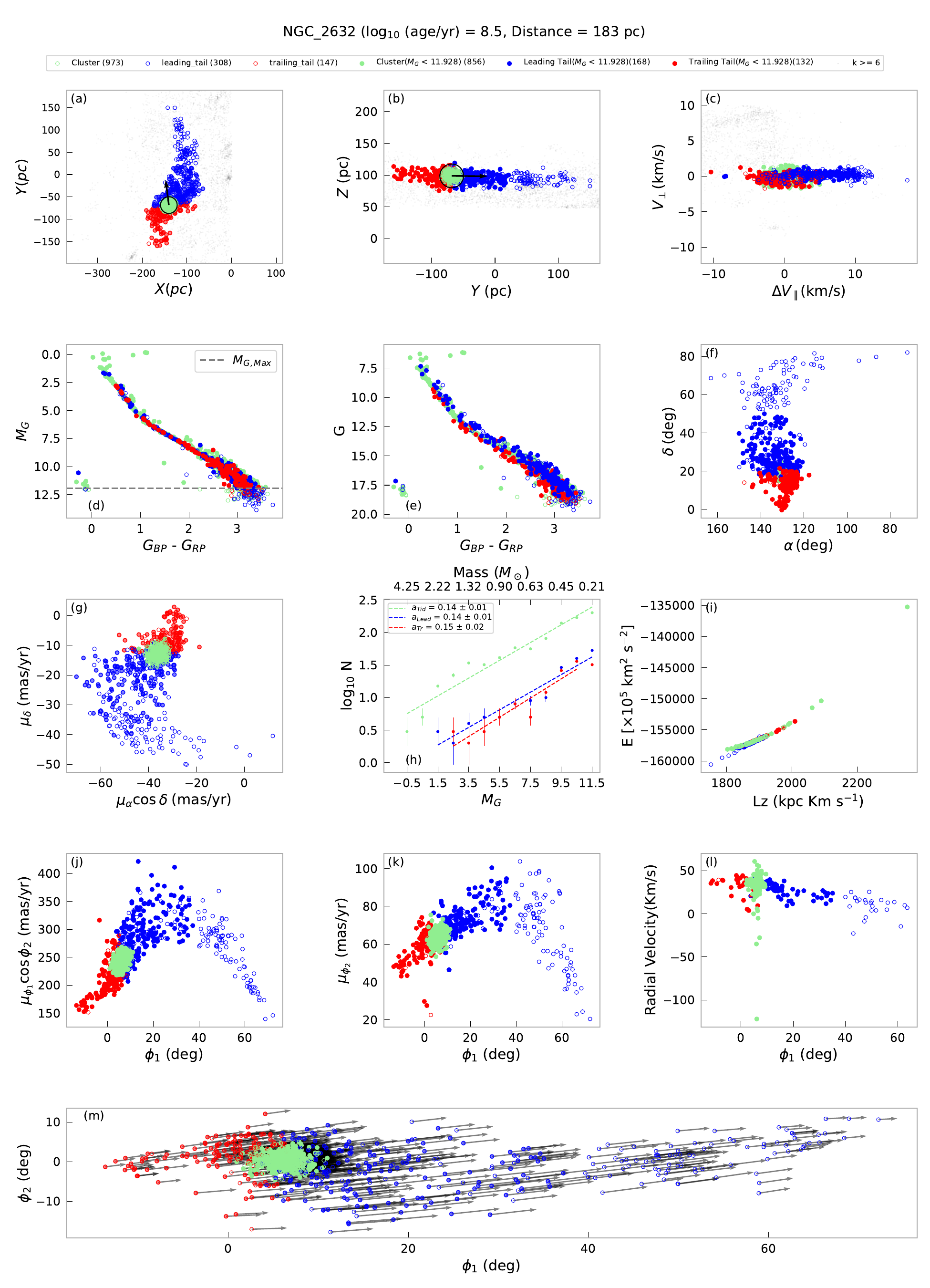}
    \caption{Same as the Figure~\ref{fig:ASCC_101} for NGC 2632}
    \label{fig:NGC_2632}
\end{figure*}

\begin{figure*}[h]
    \includegraphics[width=0.95\textwidth]{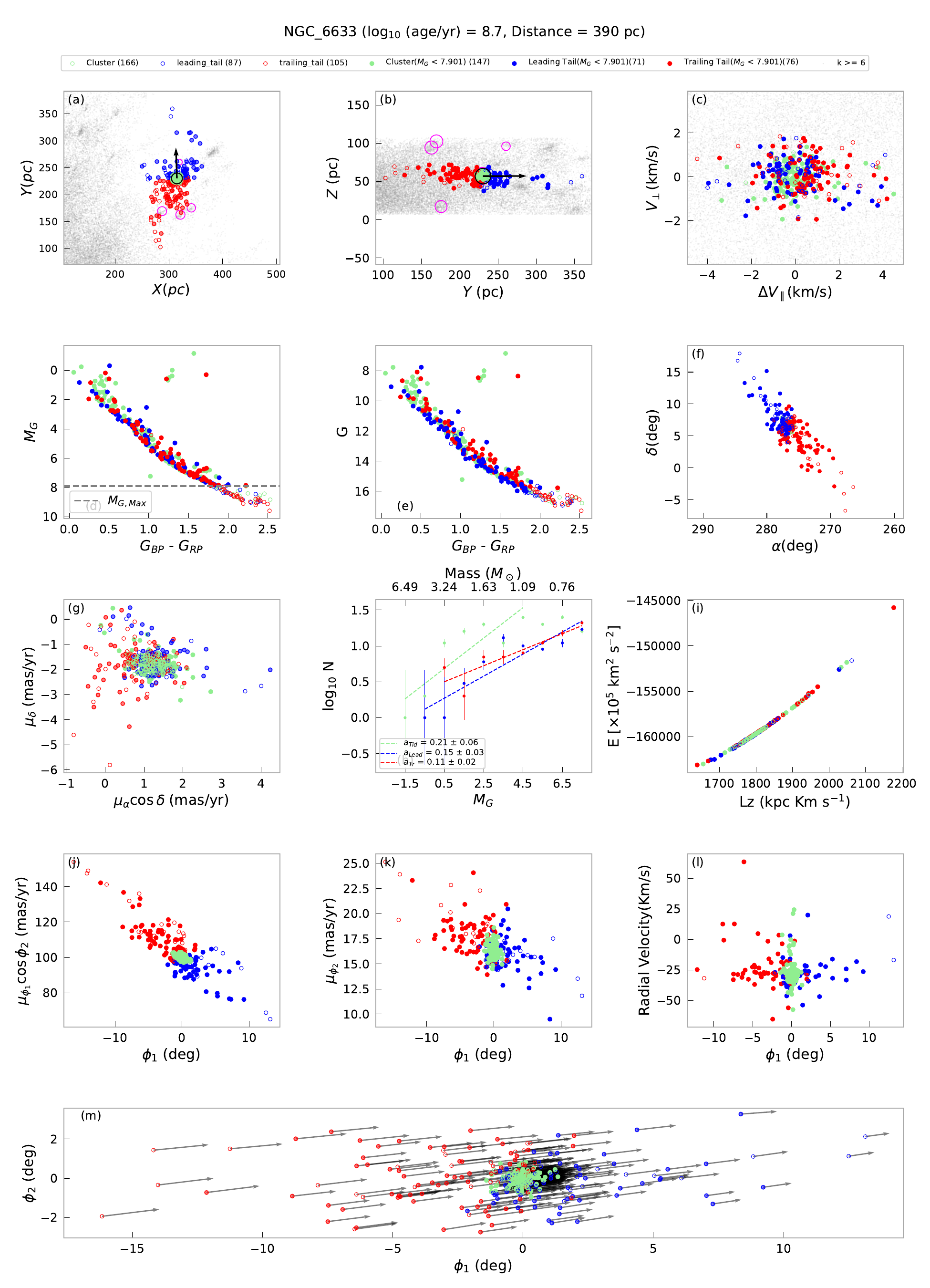}
    \caption{Same as the Figure~\ref{fig:ASCC_101} for NGC 6633}
    \label{fig:NGC_6633}
\end{figure*}

\begin{figure*}[h]
    \includegraphics[width=0.95\textwidth] {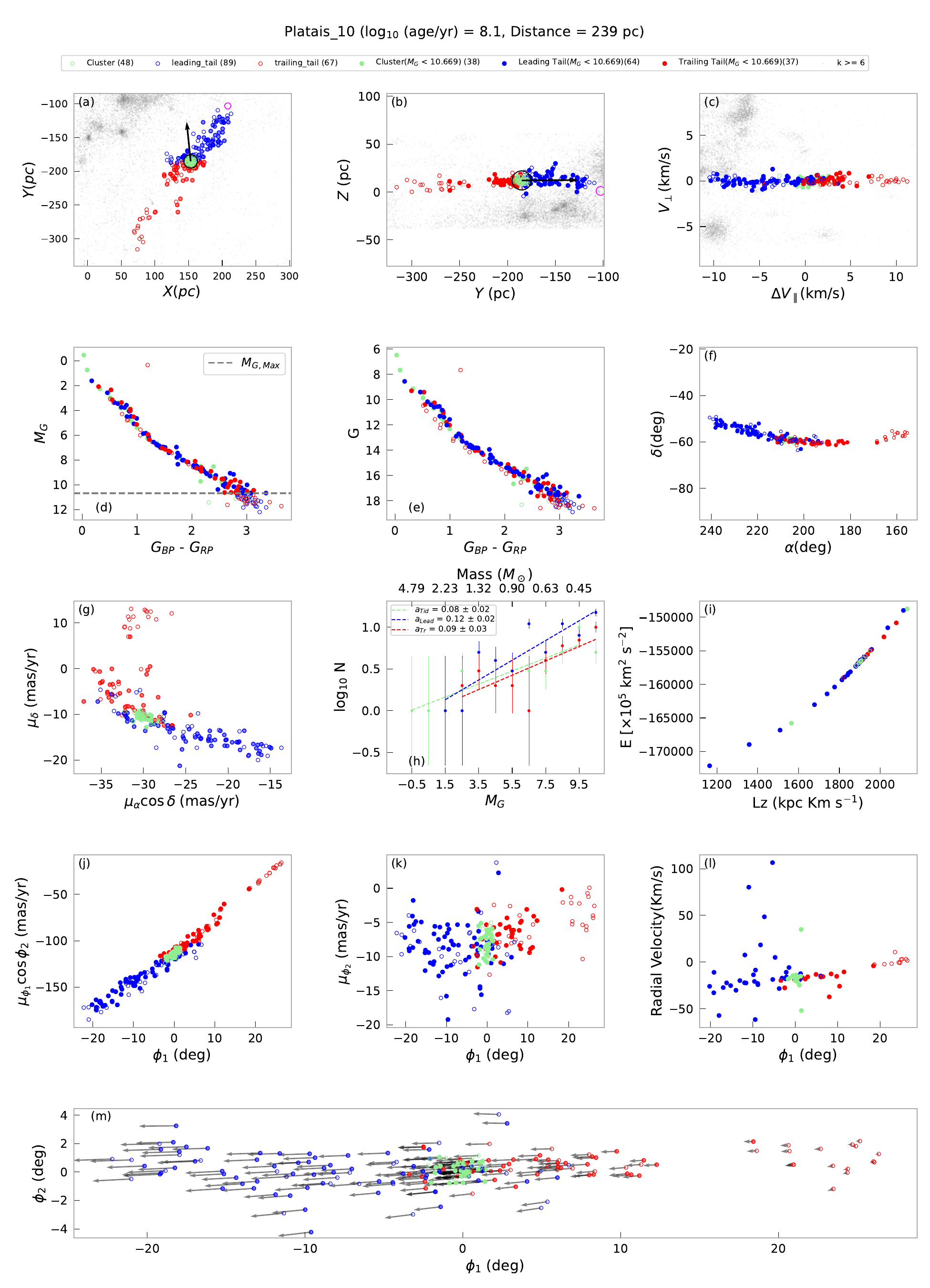}
    \caption{Same as the Figure~\ref{fig:ASCC_101} for Platais 10}
    \label{fig:Platais_10}
\end{figure*}

\begin{figure*}[h]
    \includegraphics[width=0.95\textwidth]{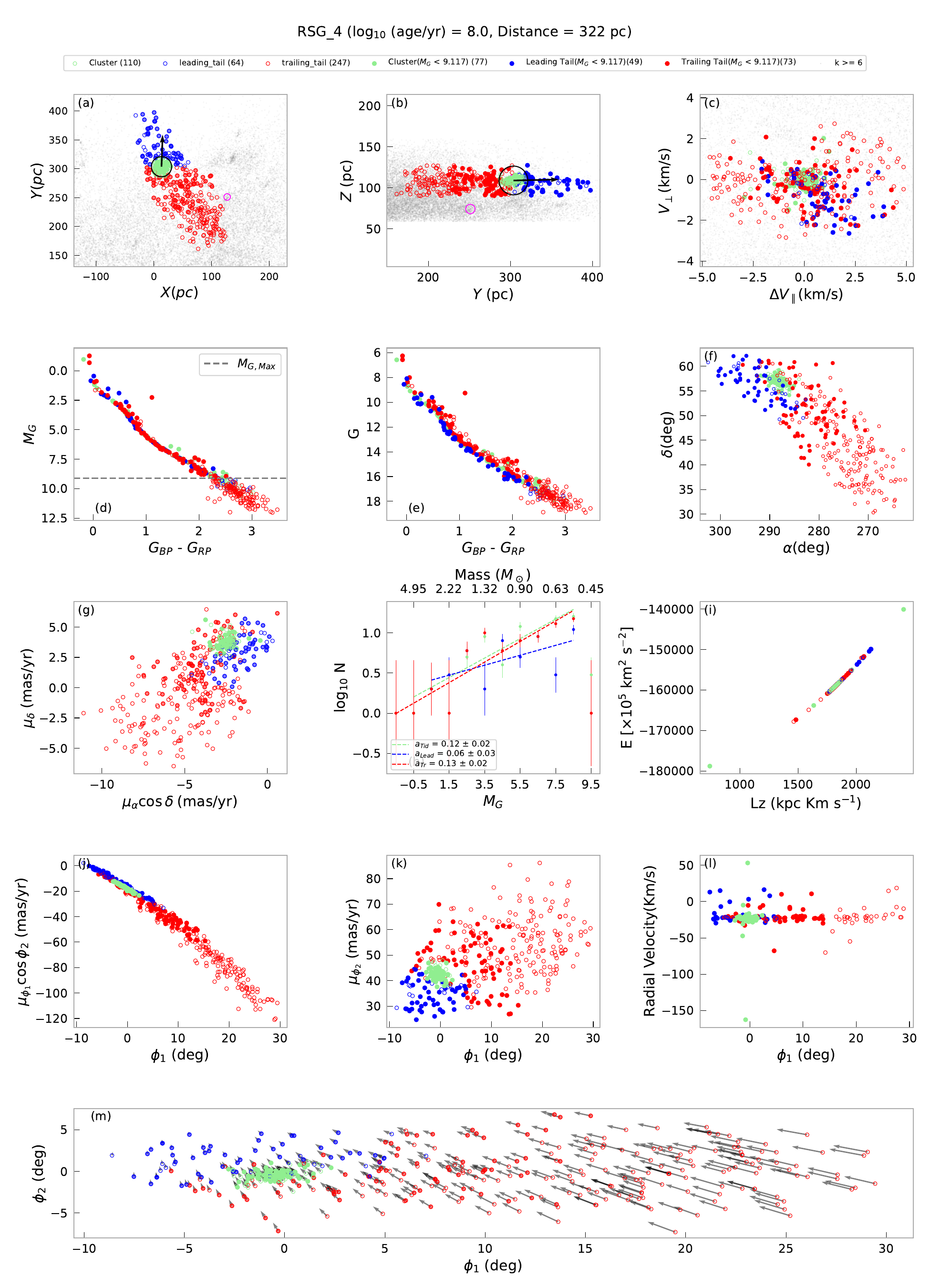}
    \caption{Same as the Figure~\ref{fig:ASCC_101} for RSG 4}
    \label{fig:RSG_4}
\end{figure*}

\begin{figure*}[h]
    \includegraphics[width=0.95\textwidth]{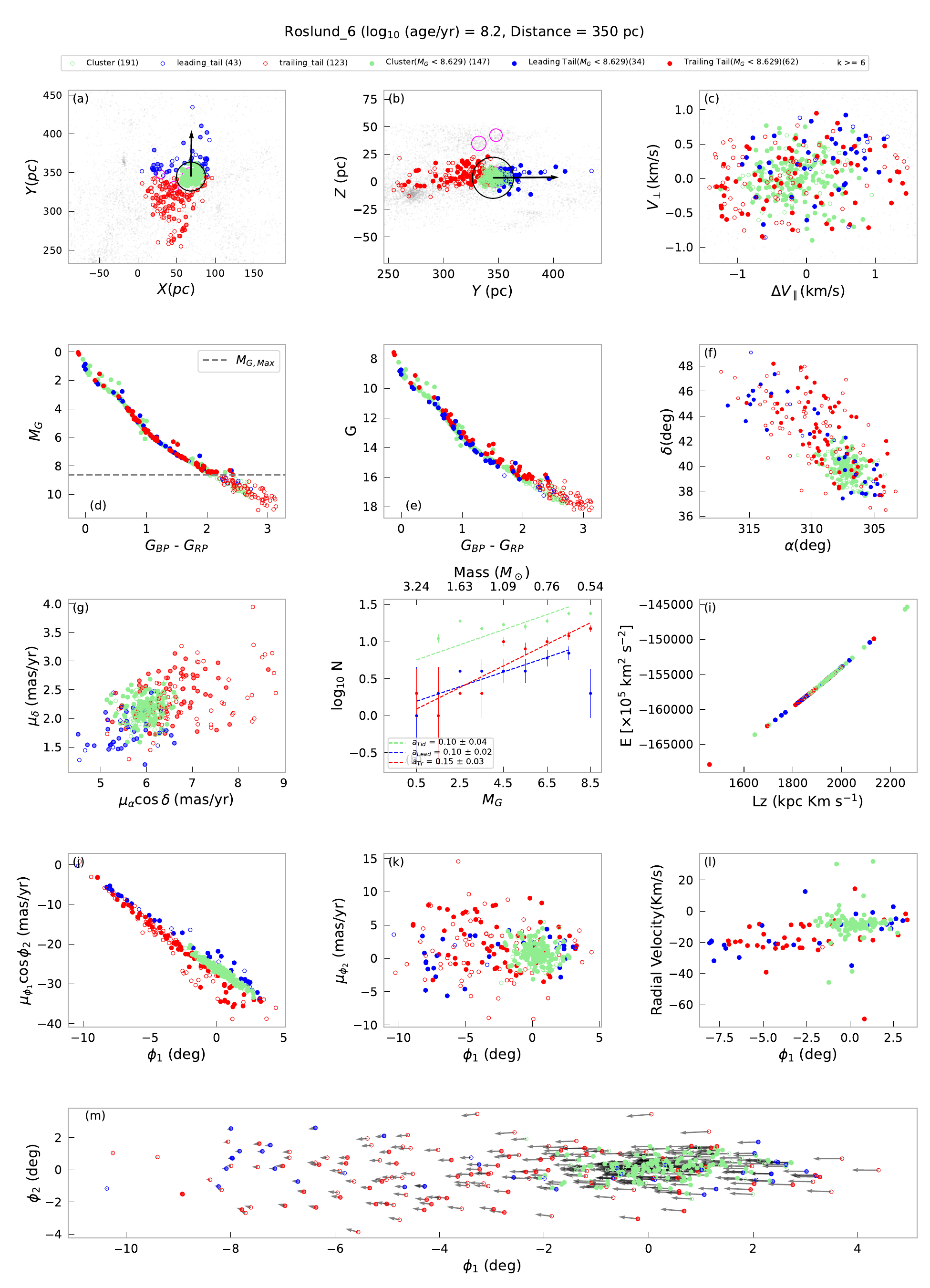}
    \caption{Same as the Figure~\ref{fig:ASCC_101} for Roslund 6}
    \label{fig:Roslund_6}
\end{figure*}

\begin{figure*}[h]
    \includegraphics[width=0.95\textwidth]{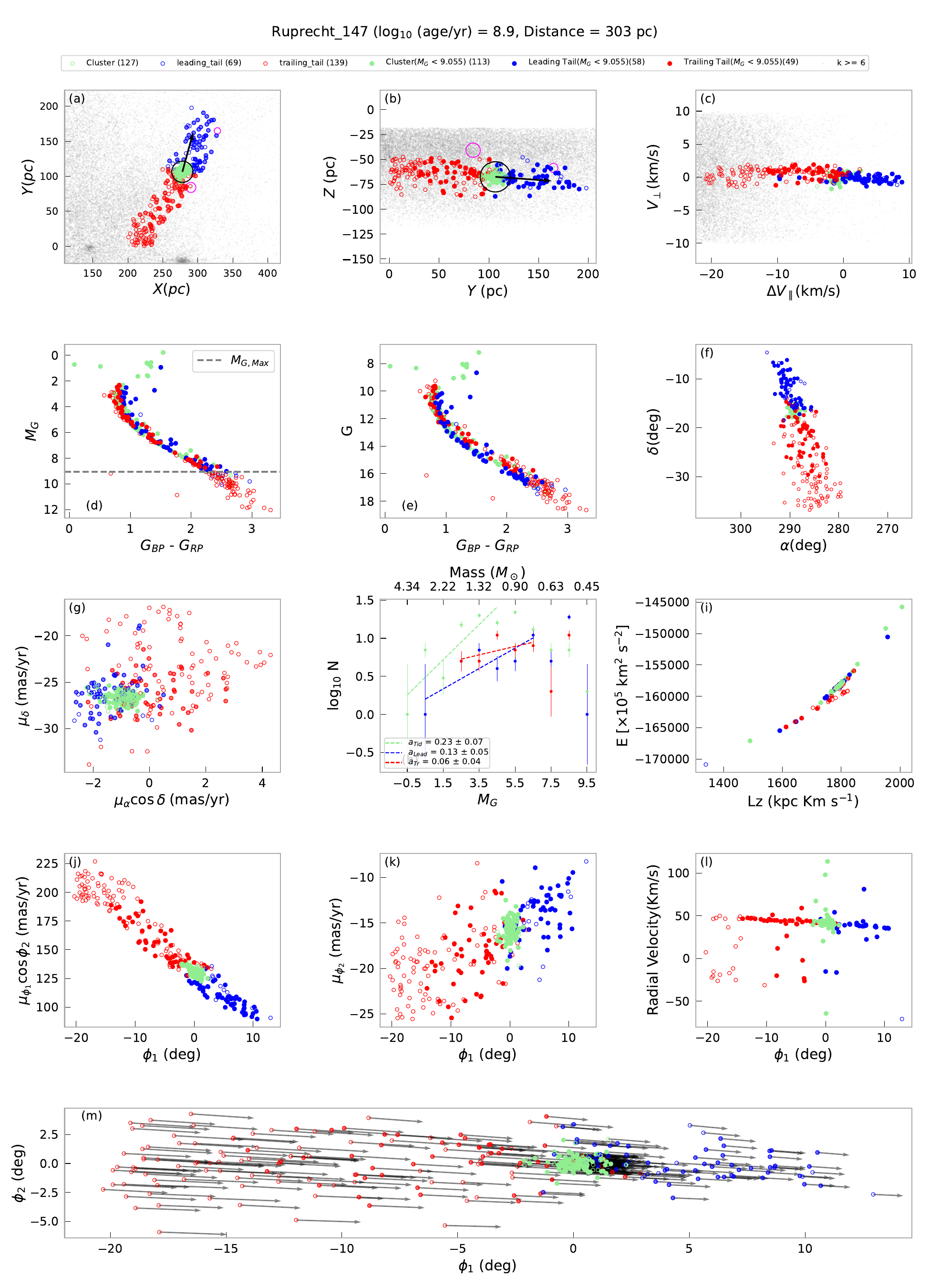}
    \caption{Same as the Figure~\ref{fig:ASCC_101} for Ruprecht 147}
    \label{fig:Ruprecht_147}
\end{figure*}

\begin{figure*}[h]
    \includegraphics[width=0.95\textwidth]{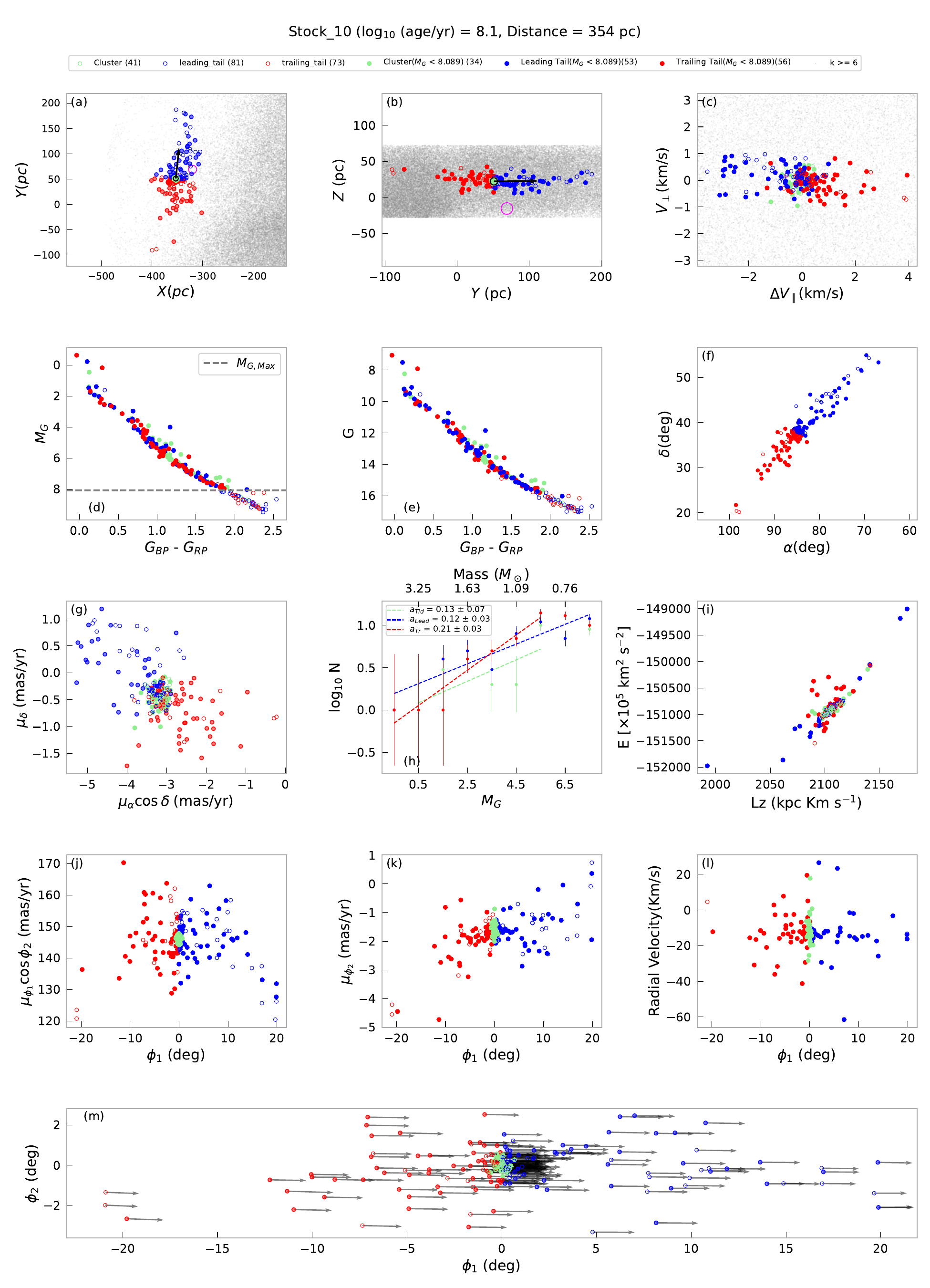}
    \caption{Same as the Figure~\ref{fig:ASCC_101} for Stock 10}
    \label{fig:Stock_10}
\end{figure*}

\begin{figure*}[h]
    \includegraphics[width=0.95\textwidth]{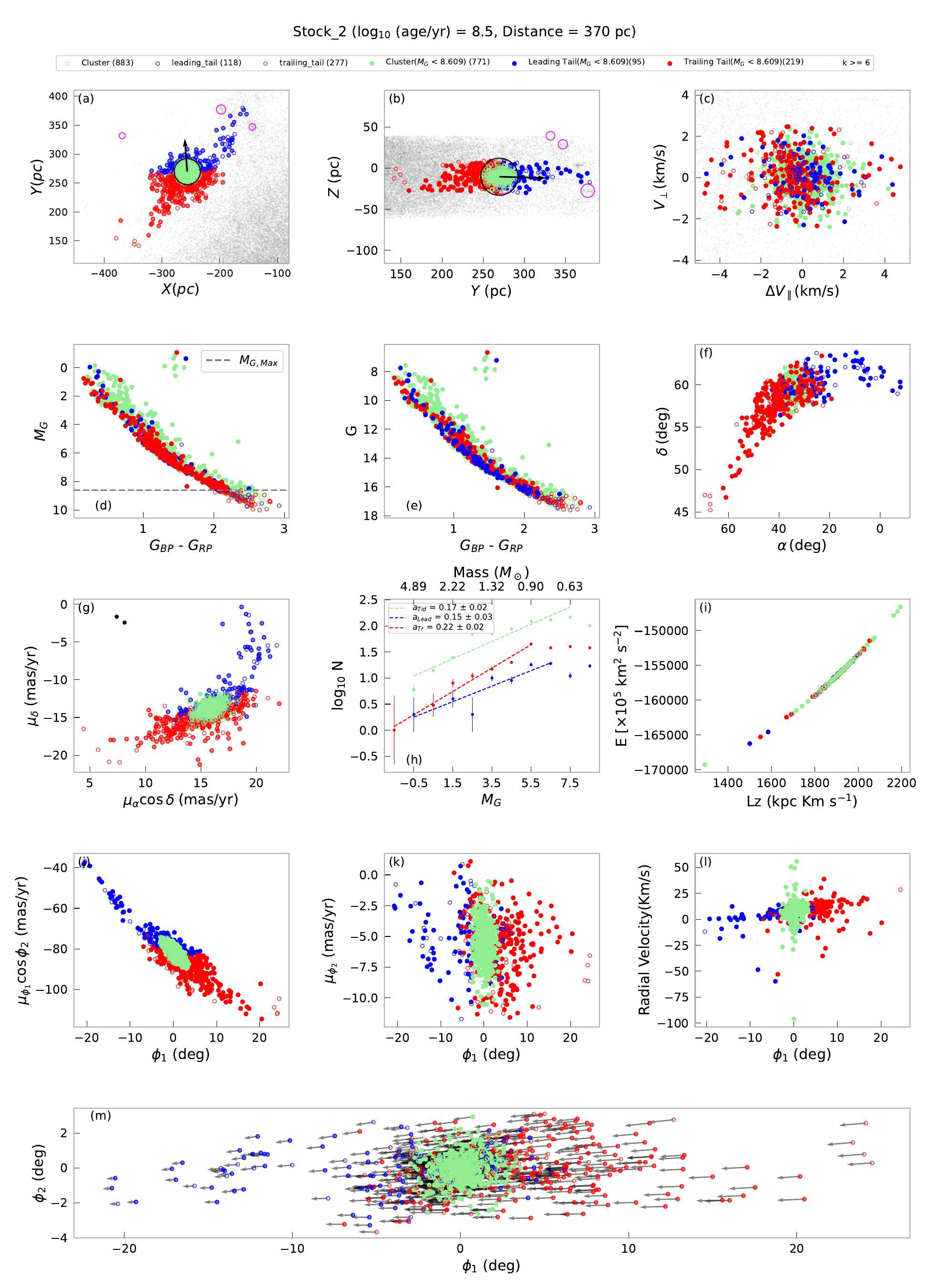}
    \caption{Same as the Figure~\ref{fig:ASCC_101} for Stock 2}
    \label{fig:Stock_2}
\end{figure*}

\begin{figure*}[h]
    \includegraphics[width=0.95\textwidth]{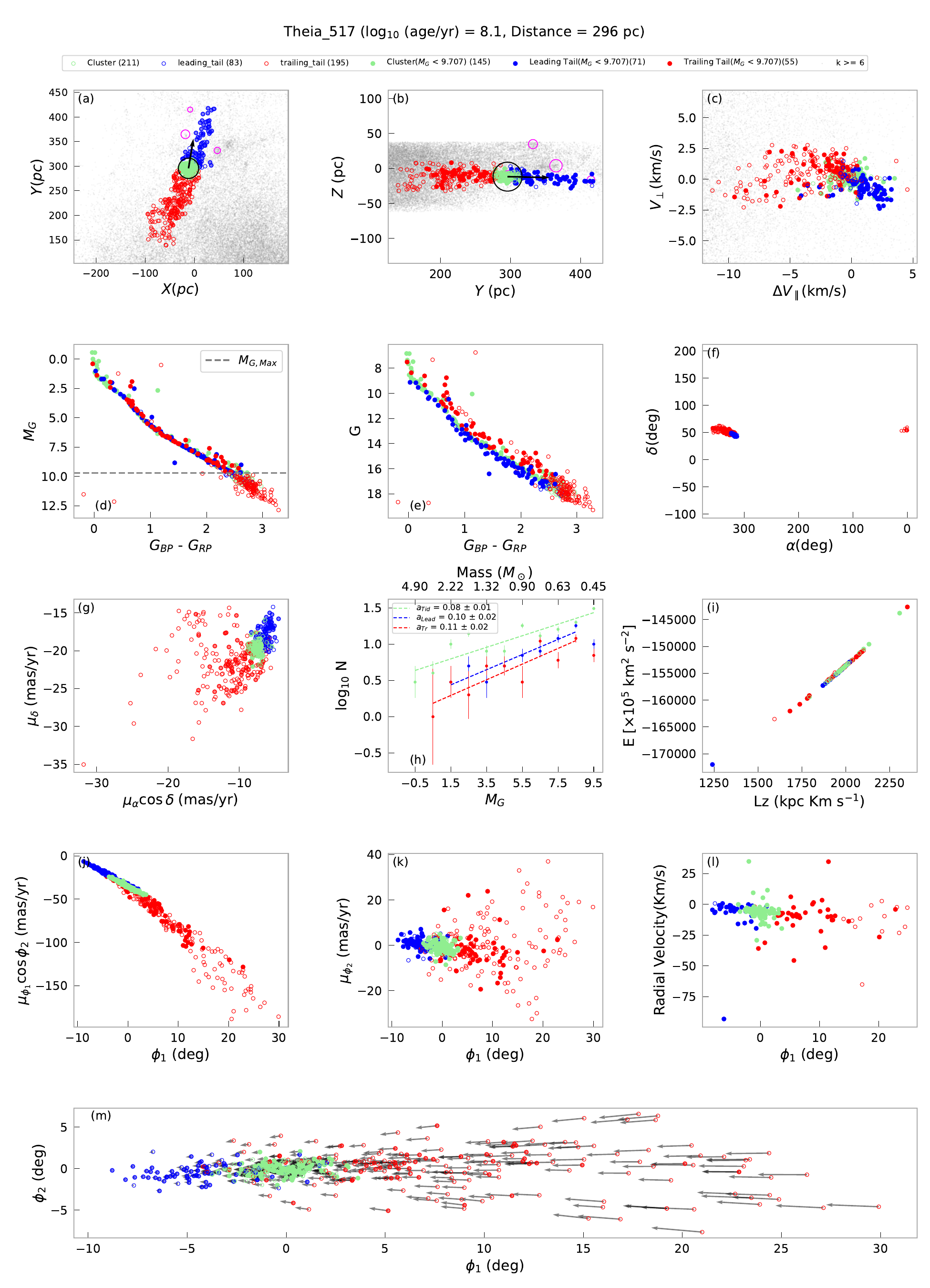}
    \caption{Same as the Figure~\ref{fig:ASCC_101} for Theia 517}
    \label{fig:Theia_517}
\end{figure*}

\begin{figure*}[h]
    \includegraphics[width=0.95\textwidth]{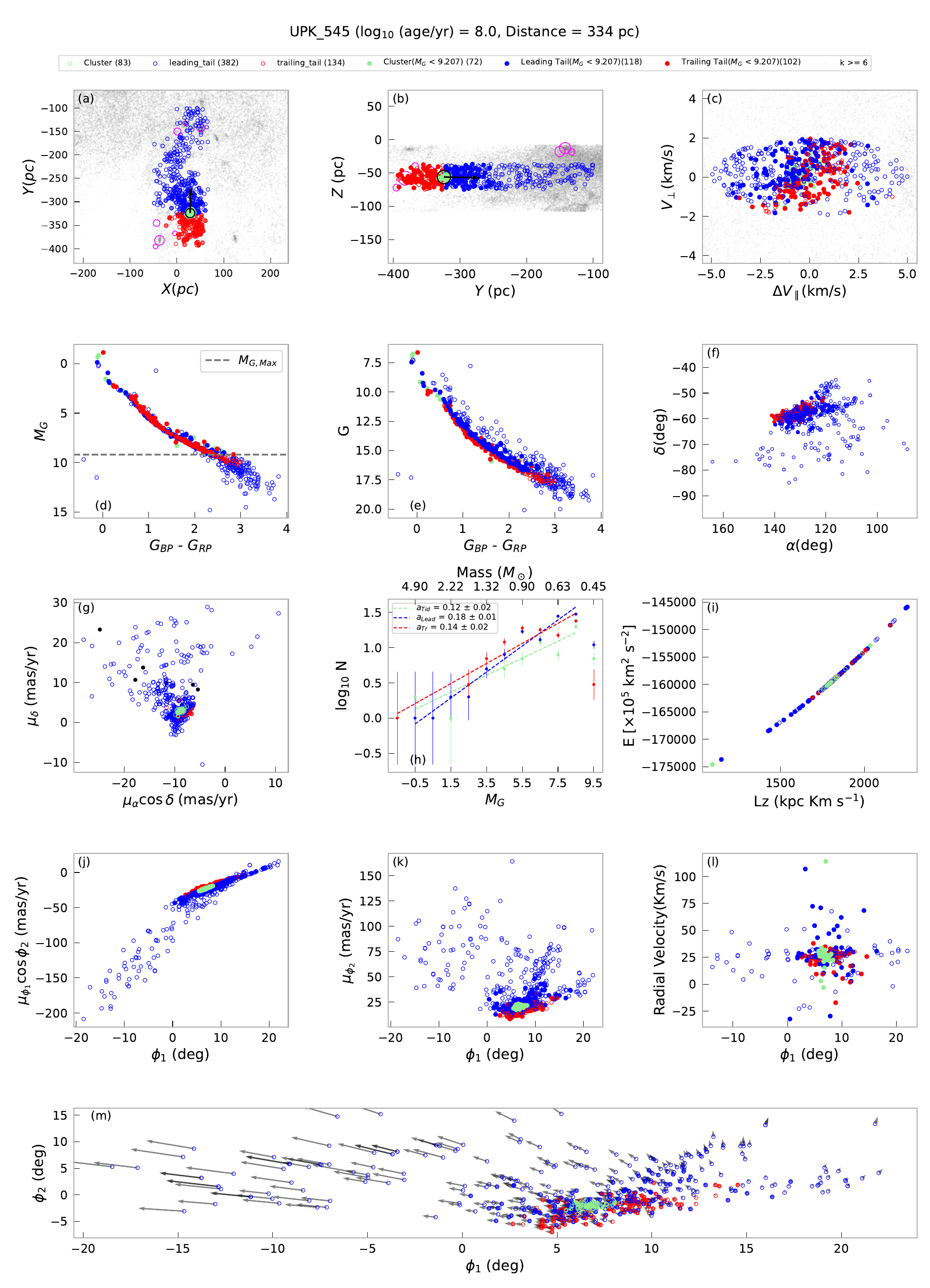}
    \caption{Same as the Figure~\ref{fig:ASCC_101} for UPK 545}
    \label{fig:UPK_545}
\end{figure*}

The caption to the appendix Figs. below~\ref{fig:ASCC_101} to ~\ref{fig:UPK_545} to is as follows:

Member stars within the tidal radius, leading tail and the trailing tail are shown as hollow circles with the light-green, blue and red colour whereas over-plotted filled circles with same colours are the corrected sample (see Section~\ref{sec: Corrected Samples}). Grey points are field stars with $k$ $\geq$ 6 in the selected phase space (see Section~\ref{sec: CP}). The panels are as follows: (a) Plot in solar system barycentric Galactic Cartesian $X-Y$ coordinates where the Sun is at (0,0) and the Galactic centre is towards the right. Black circle is the tidal radius and the black arrow is the mean motion of the cluster. The pink circles in (a) and (b) are nearby, phase space neighbouring clusters or moving groups from \citetalias{2024arXiv240305143H}. (b) Spatial distribution in $Y-Z$ coordinates. Markers similar to panel (a). (c) Convergent-point velocities. (d) Absolute CMD with dotted black line showing the magnitude cut-off ($M_G$) mentioned in Eq~\ref{eq:MGlim}. (e) Apparent CMD. (f) Spatial distribution in equatorial coordinates. (g) Vector point diagram. (k) LF (see Section~\ref{sec: LF}) for the cluster members (light-green dots), leading tail (red dots) and trailing tail (blue dots) plotted using corrected sample. The mass corresponding to the G-band magnitude is shown at top based on a solar metallicity PARSEC isochrone \citep{2012MNRAS.427..127Bressan_Parsec_iso}. The least squares straight line fits shown in the dotted lines of respective colours. (i) ($L_Z - E$) phase space (see Section~\ref{sec: LzE_test}). (j) $\phi_1$ Vs $\mu_{\phi_{1}} cos \phi_2 $. (k) $\phi_1$ Vs $\mu_{\phi_{2}} $. (l) Plot in $\phi_1$ Vs Radial Velocity (available for only the brightest stars in the Gaia data). (m) Plot in tail-aligned $\phi_1$-$\phi_2$ sky coordinates with the tail-aligned proper motions ($\mu_{\phi_{1}} cos \phi_2 $, $\mu_{\phi_{2}} $) in black arrows. Coordinates are in degree and proper motions in mas yr$^{-1}$.

\begin{comment}

Similar plots for all the clusters in the Appendix Table~\ref{Table: Tidal tails} can be found  \href{https://doi.org/10.5281/zenodo.14651597}{here}. 

\end{comment}

\newpage

% \restoregeometry

% \newpage

% \begin{figure*}
%     \centering
%     \includegraphics[width=0.95\textwidth]{figures/clones_all_xy.jpg}
%     \caption{$X-Y$ plot of all clones.}
%     \label{fig:clones_xy}
% \end{figure*}

\begin{figure*}[h]
    \includegraphics[width=0.95\textwidth]{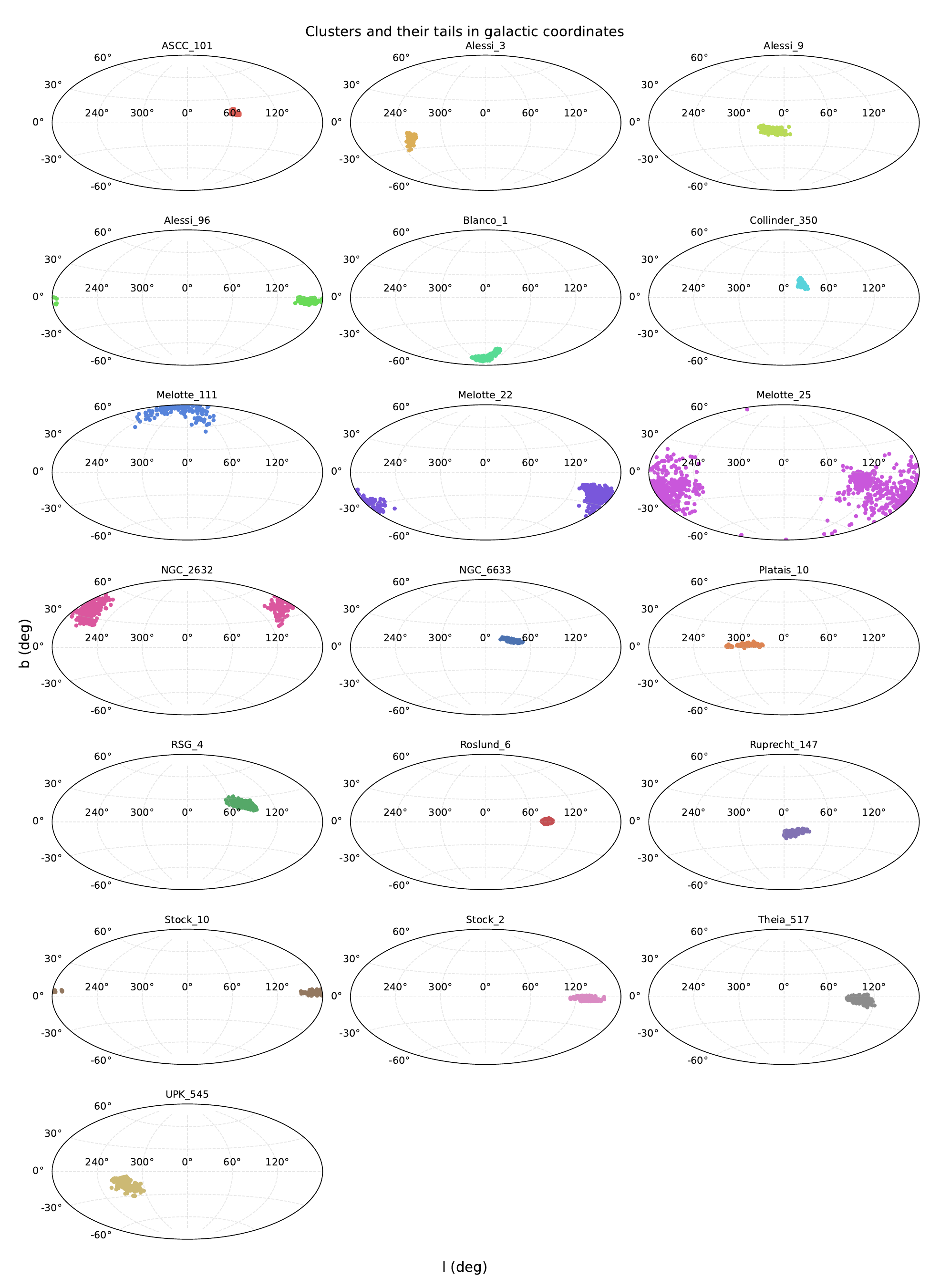}
    \caption{Distribution of cluster and their tail stars in galactic coordinates. }
    \label{fig:lb}
\end{figure*}

\end{appendix}

\end{document}